\pdfoutput=1

\let\oldcline\cline  
\let\cline\relax     

\documentclass[sn-mathphys-num]{sn-jnl}

\let\cline\oldcline  

\usepackage{graphicx}%
\usepackage{multirow}%
\usepackage{amsmath,amssymb,amsfonts}%
\usepackage{amsthm}%
\usepackage{mathrsfs}%
\usepackage[title]{appendix}%
\usepackage{xcolor}%
\usepackage{textcomp}%
\usepackage{manyfoot}%
\usepackage{booktabs}%
\usepackage{algorithm}%
\usepackage{algorithmicx}%
\usepackage{algpseudocode}%
\usepackage{listings}%

\usepackage{multirow}
\usepackage{multicol}
\usepackage{xcolor}
\usepackage[most]{tcolorbox} 
\usepackage{diagbox}
\usepackage{tabularx}
\usepackage{subcaption}
\usepackage{colortbl}
\usepackage{array}
\usepackage{rotating}

\newcommand{\ncf}{{\ensuremath{\sf NCF}}}
\newcommand{\ncc}{{\ensuremath{\sf NCC}}}
\newcommand{\rcf}{{\ensuremath{\sf rcf}}}
\newcommand{\rcc}{{\ensuremath{\sf rcc}}}

\newcommand{\nca}{{\ensuremath{\sf NCA}}}
\newcommand{\NCG}{{\ensuremath{\sf NCG}}}
\newcommand{\rca}{{\ensuremath{\sf rca}}}
\newcommand{\rcg}{{\ensuremath{\sf rcg}}}

\newcommand{\CI}{{\ensuremath{\sf CI}}}

\newcommand{\Sim}{{\ensuremath{\sf Sim}}}

\begin{document}

\title[Article Title]{Unveiling Code Clone Patterns in Open Source VR Software: An Empirical Study}


\author[1,2]{\fnm{Huashan} \sur{Chen}}\email{chenhuashan@iie.ac.cn}

\author[1,2]{\fnm{Zisheng} \sur{Huang}}\email{huangzisheng@iie.ac.cn}

\author[1,2]{\fnm{Yifan} \sur{Xu}}\email{xuyifan@iie.ac.cn}

\author[1,2]{\fnm{Wenjie} \sur{Huang}}\email{huangwenjie@iie.ac.cn}

\author[3]{\fnm{Xuheng} \sur{Wang}}\email{22331104@bjtu.edu.cn}

\author[4]{\fnm{Jinfu} \sur{Chen}}\email{jinfuchen@whu.edu.cn}

\author[5]{\fnm{Haotang} \sur{Li}}\email{haotangl@arizona.edu}

\author[6]{\fnm{Kebin} \sur{Peng}}\email{pengk24@ecu.edu}

\author*[1,2]{\fnm{Feng} \sur{Liu}}\email{liufeng@iie.ac.cn}

\author[4]{\fnm{Sen} \sur{He}}\email{senhe@arizona.edu}

\affil[1]{\orgdiv{Institute of Information Engineering}, \orgname{Chinese Academy of Sciences}, \orgaddress{\city{Beijing}, \country{China}}}

\affil[2]{\orgdiv{School of Cyber Security}, \orgname{University of Chinese Academy of Sciences}, \orgaddress{\city{Beijing}, \country{China}}}

\affil[3]{\orgdiv{School of Computer Science and Technology}, \orgname{Beijing Jiaotong University}, \orgaddress{\city{Beijing}, \country{China}}}

\affil[4]{\orgdiv{School of Computer Science}, \orgname{Wuhan University}, \orgaddress{\city{Wuhan}, \country{China}}}

\affil[5]{\orgdiv{Department of Electrical and Computer Engineering}, \orgname{University of Arizona}, \orgaddress{\city{Tucson}, \country{USA}}}

\affil[6]{\orgdiv{Department of Computer Science}, \orgname{East Carolina University}, \orgaddress{\city{Greenville}, \country{USA}}}


\abstract{Code cloning is frequently observed in software development, often leading to a variety of maintenance and security issues. While substantial research has been conducted on code cloning in traditional software, to the best of my knowledge, there is a lack of studies on cloning in {virtual reality (VR)} software that consider its unique nature, particularly the presence of numerous serialized files in conjunction with the source code. In this paper, we conduct the first large-scale quantitative empirical analysis of software clones in 345 open-source VR projects, using the NiCad detector for source code clone detection and large language models (LLMs) for identifying serialized file clones. Our study leads to a number of insights into cloning phenomena in VR software, guided by seven carefully formulated research questions. These findings, along with their implications, are anticipated to provide useful guidance for both researchers and software developers within the VR field.
}


\maketitle

\section{Introduction}
{\color{black}In recent years, Virtual Reality (VR) has gained substantial traction, finding significant applications across diverse domains such as gaming, healthcare, education, and entertainment\cite{anthes2016state,berg2017industry,kaminska2019virtual}. An increasing number of applications are being created and distributed on leading app platforms (e.g., Google Play, Apple Store, Oculus), collectively reaching approximately 200 million downloads worldwide \cite{VirtualR14:online}. However, research on the quality of VR software remains limited, with code clone—a common yet often overlooked practice in software development—being particularly unexplored within this context.

Code clone generally refers to the repetition of identical or similar code snippets in multiple places within a software project, which sometimes may pose critical challenges for software security\cite{juergens2009code}, maintainability\cite{chatterji2013effects} and changeability\cite{lozano2008assessing}. 
While there has been extensive research on the issue of code cloning in traditional software\cite{duala2008clonetracker,rajakumari2020towards,mondal2023granularity,islam2017security,solanki2016comparative,kamiya2002ccfinder,white2016deep,cordy2011nicad,feng2020nicad+}, to the best of my knowledge, our prior work\cite{huang2024study} is the first quantitative empirical study of code cloning in open-source VR software. Due to space limitations, the study \cite{huang2024study} limits its scope to source-code level code cloning. Nonetheless, VR software exhibits a unique characteristic compared to traditional software, that is, it incorporates a large number of serialized files, which are pivotal for data representation and storage in VR projects. Cloning of these files may also introduce various issues and potential risks. For instance, in cases where identical small balls are created by copying the same \verb|.prefab|, any updates (e.g., to size or material) must be applied to each copy manually, leading to potential inconsistencies. Additionally, flaws in the original \verb|.prefab|, like missing colliders or script errors, will affect all copies, making debugging more difficult. This reinforces the need for clone detection in such non-traditional code files in tandem with source code, a topic that is still surprisingly under-researched.

{
{In this study, we substantially extend our previous work\cite{huang2024study} through three key advancements: (i) we pioneer the exploration of clone-related issues in serialized files of VR software, a distinctive feature of VR systems compared to conventional software, whereas the prior study only examined source code clones. This is, to the best of our knowledge, the first work to analyze cloning in both source code and serialized files within VR software; (ii) we restructure the research framework by refining the original research questions (RQs) and evaluation metrics, and by formulating new VR-oriented questions, with the goal of shedding light on the mechanisms underlying code cloning in VR software and how these differ fundamentally from those observed in non-VR contexts; (iii) we considerably enlarge the dataset from 83 to 345 open-source VR projects, enhancing the robustness and generalizability of the findings.}

Our study uncovers a number of insights into cloning phenomena in VR software and 
highlights the unique challenges of VR software development, offering practical guidance for researchers, developers,
and industry practitioner within the VR field. Some of our insights are highlighted as follows: {(i) Both source code cloning and serialized file cloning in VR software show a consistent pattern, with larger projects generally exhibiting more clones. (ii) Code and file cloning issues are most prevalent in gaming and education software, with gaming being the most significantly impacted, highlighting the need for targeted attention in this domain. (iii) Unlike traditional software, most VR projects exhibit significantly lower source code cloning due to their heavy reliance on serialized asset files. (iv) While C\# is the most commonly used programming language in open-source VR projects, C-based projects exhibit a higher proportion of code cloning. (v) The introduction of third-party libraries in VR applications often contributes to source code cloning issues, sometimes constituting the majority of the cloning results. (vi) Inter-version code cloning is more common than intra-version code cloning, introducing significant security risks due to the potential propagation of vulnerabilities across multiple versions of the software. (vii) The cloning behavior of asset files varies with complexity. Complex files like scenes have lower clone levels, simpler ones like materials have higher,
and prefab files fluctuate with development style.}

To sum up,  this paper provides the following main contributions:

\begin{itemize}

    \item We highlight the unique nature of code cloning in VR software relative to traditional software and conduct a thorough quantitative empirical analysis of both source code clones and serialized asset file clones in our constructed dataset of 345 open-source VR projects.

    \item We propose a detection method based on large language models to identify clones in the unique serialized asset files of VR software, presenting a novel approach to code clone detection in this domain.

    \item We derive several insights from the experimental results that deepen our understanding of VR software, providing practical guidance for researchers, developers, and industry practitioner in the VR domain.
\end{itemize}
}

\noindent\textbf{Paper Organization.} The paper is structured as follows. Section \ref{sec:background} provides the background of VR software. Section \ref{sec:methodology} explains our methodology. Section \ref{sec:evaluation}  presents the results of seven research questions. Section \ref{sec:discussion} suggests future work. Section \ref{sec:related_work} discusses related work. Section \ref{sec:conclusion} concludes the paper.

\section{Background}
\label{sec:background}

In this section, we provide a brief introduction to VR software and discuss methods for detecting code cloning issues within it.

\subsection{VR Software Architecture}
\label{sec:vr}
{\color{black}VR software tightly integrates source code with serialized asset files, collectively shaping the application's functionality and immersive features. During runtime, source code engages directly with assets to perform essential tasks, such as loading and rendering 3D models, triggering and managing animations, and processing user inputs to deliver interactive experiences. In the realm of VR software development, platforms like Unity \cite{unity_manual,goldstone_unity,grayson_unity_vr} play a pivotal role, offering a comprehensive architectural model that synthesizes code-directed operational behaviors with immersive, asset-structured environmental elements. 

\textbf{Source Code.} 
In VR software, the source code primarily consists of scripts that define the core functionality of the application. Typically written in C\#, these scripts govern object behavior, manage user interactions, and implement key features such as rendering, physics simulations, and input handling. In Unity, source code adopts a content-logic decoupling and component-based architecture, where individual scripts are assigned to game objects to regulate specific aspects of their functionality. While this design enhances reusability, it can also result in redundancy, particularly in large-scale or collaborative development contexts.

\textbf{Serialized Asset Files.} 
In VR software, asset files serve as the backbone for delivering visual, auditory, and interactive content. In Unity, assets are serialized for efficient storage and retrieval.  Common serialized files in VR software include asset files such as material files (\verb|.mat|, \verb|.shader|, \verb|.cginc|), scene files (\verb|.unity|, \verb|.scene|, \verb|.navmesh|), resource files (\verb|.fbx|, \verb|.wav|, \verb|.png|), and template files (\verb|.prefab|), along with configuration files (\verb|.json|, \verb|.ini|) and metadata files (\verb|.meta|) bound one-to-one with asset files \cite{unity_manual}. While the serialized structure ensures platform interoperability and smooth resource integration, its component-based design can result in redundancy across projects or scenes.

\subsection{VR Code Clone}
\label{sec:vr_code_clone}
In light of the aforementioned VR software architecture, this paper extends the concept of code clone beyond its conventional scope at the source code level to include the cloning of data embedded in serialized asset files.

\subsubsection{Source Code Clone}
\label{sec:source_code_clone}

In the context of software development, source code clone denotes the duplication of similar code fragments within a codebase, where code fragment refers to a continuous segment of source code. Generally, there are four clone forms\cite{roy2009comparison,rattan2013software,white2016deep,ain2019systematic,7886988}: (i) {\em identical clone}, referring to identical code fragments except for variations in comments and layout; (ii){\em lexical clone}, referring to changes in identifier names and lexical values on the basis of the identical clone; (iii){\em syntactic clone}, referring to syntactically similar code fragments that add statements of mutual addition, modification, or deletion on the basis of the identical and lexical clone; (iv){\em semantic clone}, referring to syntactically different code fragments that implement the same functions. Note that the first three clone types indicate textual similarity whereas the last type reflects functional similarity.

Various approaches for source code clone detection have been proposed in the literature,
as elaborated below. (i)Textual-based clone detection \cite{Banda2015CodeCD} compares the raw text of code to identify clones, typically relying on string-matching techniques. This approach is suitable for detecting explicit and simple clones, such as exact copy-pasted code. Representative implementations of this method include Simian \cite{simian} and Duplo \cite{duplo}. (ii)Token-based clone detection \cite{Sheneamer2016ASO,6726700} transforms the source code into tokens (lexical units) before comparison. By ignoring non-semantic elements such as whitespace, comments, and optionally identifiers, it identifies clones based on token sequences. Representative implementations of this method include NiCad \cite{NiCadClo0:online,cordy2011nicad} and SourcererCC \cite{7886988}.
(iii)Syntax-based \cite{8719895,zakerinasrabadi2023systematicliteraturereviewsource} clone detection analyzes the abstract syntax tree (AST) of source code to compare the syntactic structure of code fragments, allowing it to eliminate formatting differences and detect syntactically similar code. Representative implementations of this method include CloneDR \cite{CloneDR:online,OCallahan2003CloneDR} and Deckard \cite{jiang2007deckard}.
(iv)Semantic-based clone detection \cite{SHENEAMER2018405,Zhang_2023} identifies clones by analyzing the semantic structure of code fragments, focusing on whether the code fragments are functionally equivalent rather than merely syntactically similar. Representative implementations of this method include SCAM \cite{jiang2008scam}.
(v)Learning-based clone detection \cite{10.1145/3381307.3381310,zakerinasrabadi2023systematicliteraturereviewsource} leverages machine learning or deep learning techniques to train models that automatically identify code clones. Recently, a rising number of studies have explored the use of Large Language Models (LLMs) for detecting code clones. \cite{majdinasab2024trained}. 
Representative implementations of this method include CodeBERT \cite{CodeBERT:online,Feng2020CodeBERT}, BigCloneBench \cite{svajlenko2015bigclonebench} and CodeGPTSensor \cite{xu2024distinguishing}. 

{It is worth noting that the first four detection methods are capable of detecting identical clones, lexical clones, and certain types of syntactic clones. In contrast, learning-based methods, including those utilizing LLMs, excel at identifying partial semantic clones but are less effective in detecting the other three types of clones compared to traditional approaches\cite{dou2023understandingcapabilitylargelanguage,majdinasab2024trained,kabore2024cross,khajezade2024efficacy}. Considering these facts, this study intends to employ Nicard, the state-of-the-art technique within traditional methods, for source code clone detection.}

\subsubsection{Serialized File Clone}
\label{sec:asset_files_clone}

Recall that serialized asset files in VR software serve as data containers, systematically defining the core visual and interactive aspects of a virtual environment. The structured nature of asset files is prone to cause ``data clone", where configurations, asset references, or component settings are duplicated across files. 
Therefore, serialized asset file clone focuses on identifying duplicated or near-duplicated asset data, such as repeated material definitions, component hierarchies, or configuration settings. Existing detection techniques used for identifying data clone include: (i)Content-based methods, which typically construct feature vectors of the content for similarity comparison \cite{salton1983introduction,jones1972statistical,levenshtein1966binary,jaccard1901distribution}. Representative implementations of this method include Scikit-learn \cite{scikit-learn}.
(ii)Fingerprint-based methods, such as rolling hashes \cite{10.14778/2850469.2850470}, which maps similar content to an identical hash value, and MD5 \cite{rivest1992md5}, which detects file clones by comparing the similarity of hash values. Representative implementations of this method include LSH \cite{10.14778/2850469.2850470} and Duplicate File Finder \cite{duplicate-file-finder}.
(iii)Deep learning-based methods, such as word vector models \cite{Pennington2014GloVeGV,bojanowski2017enrichingwordvectorssubword,peters2018deepcontextualizedwordrepresentations} and Transformer-based models \cite{devlin2019bertpretrainingdeepbidirectional}, which train neural networks to identify textual similarity. Representative implementations of this method include Hugging Face Transformers \cite{hugging-face-transformers}.

Among these methods, content-based methods are simple and easy to implement, but they require significant computational resources and time, limiting their applicability to small-scale texts. Fingerprint-based methods, while computationally efficient, are sensitive to local changes and noise in the text. Deep learning-based methods, on the other hand, require substantial data and time for training, but with the advent of large models, this limitation has been significantly alleviated. In most scenarios, deep learning-based methods outperform the other approaches. 

{
The nature of serialized asset files—often semi-structured, heterogeneous, and less amenable to syntax-based analysis—makes LLMs more suitable for capturing semantic similarities in this context. The latest studies indicate that GPT-4o, owing to its powerful contextual understanding and generative capabilities, often surpasses traditional embedding-based methods and other generative models in text similarity detection tasks, particularly those requiring reasoning and handling contextual variations \cite{liu2023gpt,zhang2023gpt}. Given these facts, we propose using GPT-4o to identify clones in serialized asset files.}}

\newcommand{\modified}[1]{\textcolor{blue}{#1}}
\newcommand{\deleted}[1]{\textcolor{blue}{\st{#1}}}

\section{Methodology}
\label{sec:methodology}

The research methodology overview is shown in Figure \ref{fig:overall}.
We begin by searching for VR projects in GitHub and existing datasets. After selecting 658 projects, we filter them by programming language, reducing the dataset to 416 projects. We then manually inspect and deduplicate these projects, resulting in a final dataset of 345 unique projects. We then propose detection methods for both source code cloning and serialized asset file cloning. Next, we formulate a set of metrics for measuring VR software cloning from various perspectives. Afterward, we conduct code cloning analysis from three angles to answer seven carefully formulated research questions.

\begin{figure}[t]
    \centering
    \includegraphics[width=\linewidth]{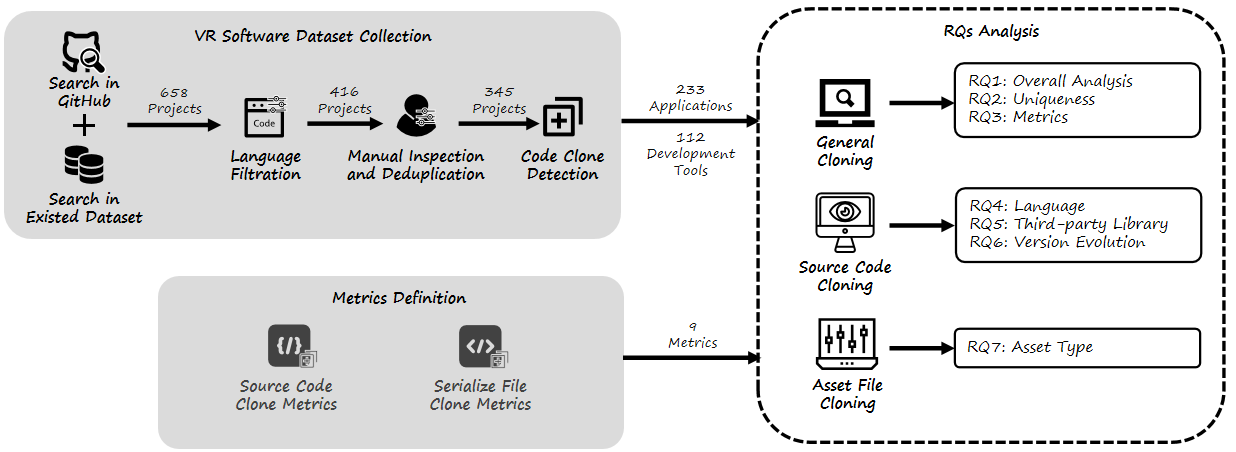}
    \caption{Overview of Our Empirical Study.}
    \label{fig:overall}
\end{figure}

\subsection{Dataset Construction}
\label{sec:dataset_construction}

{\color{black}Despite the availability of several datasets \cite{rzig2023, nusrat2021developers, rodriguez2017empirical, huang2024study} from prior studies, we have reconstructed a dataset tailored to meet four key criteria crucial for the objectives of this study: coverage to include a wide range of VR software projects, diversity to encompass varied types of VR software projects, popularity to focus on prominent GitHub repositories, and usability to enable efficient detection. The generation of the dataset follows four phases, as described below.

\textbf{Project selection.} 
Given the dynamic accessibility of open-source projects on GitHub,  we construct our dataset by combining the datasets disclosed in two latest studies \cite{rzig2023,huang2024study} and augmenting them with the latest projects published afterward. The dataset provided by \cite{rzig2023} includes 314 open-source VR software projects, comprising 164 independent projects, 63 organizational projects, and 83 academic projects, with 4 of them no longer accessible. The dataset reported by \cite{huang2024study} initially consists of 326 popular open-source VR software projects prior to any further processing. We further apply the same filtering criteria as outlined in \cite{Nusrat2021How} to incorporate the most recent data,  resulting in the inclusion of 22 additional projects to enhance the dataset. By integrating the results of the three methods, we compile a preliminary dataset consisting of 658 open-source VR software projects.

\textbf{Language filtration.} 
Considering the wide range of development environments for open-source VR projects on GitHub, we propose to identify the projects that can be analyzed by off-the-shelf clone detection tools. For this purpose, we consider NiCad \cite{cordy2011nicad}, a publicly available, state-of-the-art tool for source code clone detection, supporting languages such as C, C\#, Java, Python, PHP, Ruby, Swift, ATL, and WSDL. By applying this tool to the previously obtained dataset, we exclude projects implemented in languages not supported by NiCad, resulting in 416 projects suitable for further analysis.

\textbf{Manual inspection and deduplication.} 
To ensure dataset accuracy, we manually review the tags and ``README.md'' files of each project, filtering out non-VR applications. Subsequently, we eliminate duplicate entries within the 416 projects, producing a final dataset of 345 projects, referred to as the VR-345 dataset. The dataset comprises 233 applications and 112 development tools.

}

{\color{black}
\subsection{Code Clone Detection}
\label{sec:clone_detection}
In this section, we provide detailed approaches for detecting cloning in both source code and serialized asset files, and define a set of metrics to quantify cloning from multi-faceted perspectives.

\subsubsection{Source Code Clone Detection}\label{Source_Code_Clone_Detection}
We utilize various parameters of NiCad-6.2 to detect source code clones within the constructed dataset, followed by a detailed evaluation of the detection outcomes to compute relevant clone metrics. Recall that code clones are typically classified into four types: {\em identical}, {\em lexical}, {\em syntactic}, and {\em semantic} clones, as described in Section \ref{sec:source_code_clone}. As shown in Table \ref{tab:nicad_setting}, NiCad-6.2 is capable of detecting the first three types of clones, which are further refined into six distinct subtypes: Type1 represents identical clones; Type2 and Type2c correspond to lexical clones, differing in their handling of identifiers through two methods: blind and consistent; Type3-1, Type3-2, and Type3-2c correspond to syntactic clones, where a dissimilarity threshold, $1-\tau$, with $\tau$ denoting the similarity threshold defined above, is applied to enable the detection of near-miss clones. For example, a dissimilarity threshold of ``0.1'' allows for up to 10\% difference. We set the dissimilarity threshold to ``0.3'', which is the default value. The cloning granularity may be set at the function or block level. We opt for function-level granularity to achieve results with greater detail and specificity.

We perform source code clone detection on the VR-345 dataset. The process involves normalizing and parsing source code files for NiCad compatibility, followed by clone analysis to identify clone pairs and classify them into clone classes. For each project, we calculate key metrics, including \ncc, \ncf, \rcf, and \rcc, to evaluate the clone detection results.

\begin{table*}[t]
\caption{NiCad Settings for Source Code Clone Detection}
    \centering
    \resizebox{\textwidth}{!}{
    \begin{tabular}{|c|c|c|c|c|c|c|}
    \hline
    \multirow{2}*{\textbf{Parameters}} & \multicolumn{6}{c|}{\textbf{Target Clone Types}} \\
    \cline{2-7}
     & \textbf{Type1} & \textbf{Type2} & \textbf{Type2c} & \textbf{Type3-1} & \textbf{Type3-2} & \textbf{Type3-2c} \\
    \hline
    \textbf{Dissimilarity Threshold} & / & / & / & 0.3 & 0.3 & 0.3 \\
    \hline
    \textbf{Identifier Renaming }& none & blind & consistent & none  & blind  & consistent  \\
    \hline
    \textbf{Granularity} & function & function & function & function & function & function \\
    \hline
    \end{tabular}
    }
    \label{tab:nicad_setting}
\end{table*}

\subsubsection{Serialization File Clone Detection.}\label{Serialization_File_Clone_Detection}
\label{sec:serialization_file_detection_method}

{Unlike traditional source code clones, asset clones often contain structured metadata, binary-encoded content, and heterogeneous formats. These characteristics demand semantic-level interpretation beyond surface-level syntax, thereby increasing the complexity of clone detection. While previous studies have demonstrated GPT-4o as the most effective LLM for text similarity detection tasks, it remains necessary to assess whether this conclusion holds in the context of asset file clone detection. Given the absence of a standardized dataset for this specific task, we design a three-step validation procedure: (i) independently evaluating 200 randomly selected file pairs using three different detection models;
(ii) adopting the majority opinion (i.e., when two models agree) as a provisional benchmark in cases of conflicting results; and
(iii) performing manual verification through a consensus-based review involving three researchers. As shown in Table \ref{tab:llm_results}, the final performance metrics confirm that GPT-4o remains the most effective model for asset file clone detection in our evaluation. }

\vspace{-4mm}
\begin{table}[htbp]
\centering
\caption{Comparison of File Clone Detection Performance for Different LLMs.}
\label{tab:llm_results}
\begin{tabular}{|c|c|c|c|c|}
\hline
\textbf{Model} & \textbf{Precision(\%)} & \textbf{Recall (\%)} & \textbf{F1-score (\%)} & \textbf{Accuracy (\%)} \\
\hline
GPT-4o & 98.0 & 97.5 & 97.7 & 98.0 \\
\hline
Claude 3.5 & 95.0 & 96.0 & 95.5 & 95.5 \\ 
\hline
Gemini 1.5 pro & 94.5 & 93.0 & 93.7 & 94.0 \\
\hline
\end{tabular}

\footnotesize
Note: (i) The consistency rate of the three models: 92.5\% (185/200); (ii) GPT-4o got 11 edge cases correct.
\end{table}
}
\vspace{-4mm}

We employ a two-step Chain of Thought (CoT) \cite{10.5555/3600270.3602070} technique using the GPT-4o large language model to identify similarities in serialized files for clone detection. Unlike traditional one-step question-answering prompts, this method decomposes complex problems into smaller, sequential steps, where each step’s output informs the next. Studies \cite{fu2023complexitybasedpromptingmultistepreasoning,li2023makinglargelanguagemodels,zhang2024multimodalchainofthoughtreasoninglanguage} support the effectiveness of CoT prompting in enhancing reasoning tasks, including clone identification.

\begin{table}[htbp]
\caption{\centering{Step 1 of the CoT.}}
\centering
\begin{tabularx}{\textwidth}{|c|c|X|}
\hline
\textbf{Step} & \textbf{Types} & \multicolumn{1}{c|}{\textbf{Prompts}} \\
\hline
\multirow{8}*{1} & \multirow{1}*{Simple} & {\em Please analyze the file pairs and determine if they are clone files. }\\
\cline{2-3}
& \multirow{2}*{Structure} & {\em Please analyze the file pairs based on their component structures and determine if they are clone files. }\\
\cline{2-3}
& \multirow{2}*{Semantic} & {\em Please analyze the file pairs based on their functional semantics and determine if they are clone files.} \\
\cline{2-3}
& \multirow{2}*{Similarity} & {\em Please analyze the similarity of the file pairs and provide a similarity score between 0\% and 100\%.} \\
\hline
\end{tabularx}
\label{tab:prompts_1}
\end{table}

Step 1 of the CoT framework requires the LLM to evaluate whether a single file pair qualifies as a clone pair. To enable this process, we design four distinct prompt types: simple binary analysis, structural analysis, semantic analysis, and similarity analyses. Details of each prompt are provided in Table \ref{tab:prompts_1}. {We assess the effectiveness of these prompts by testing them on 200 randomly selected file pairs. The final performance is characterized using four standard metrics: precision, recall, F1-score, and accuracy. These metrics are derived through majority voting (using GPT-4o, Claude 3.5, Gemini 1.5 Pro), followed by manual inspection, in accordance with the three-step validation framework previously applied in LLM comparisons.}

\vspace{-4mm}
\begin{table}[htbp]
\centering
\caption{Comparison of File Clone Detection Performance for Different Prompts.}
\label{tab:prompt_results}
\begin{tabular}{|c|c|c|c|c|}
\hline
\textbf{Prompt} & \textbf{Precision(\%)} & \textbf{Recall (\%)} & \textbf{F1-score (\%)} & \textbf{Accuracy (\%)} \\
\hline
Simple & 52.5 & 40.0 & 45.0 & 45.0 \\
\hline
Structure & 60.0 & 80.0 & 68.6 & 55.0 \\ 
\hline
Semantic  & 70.0 & 60.0 & 64.6 & 65.0 \\
\hline
Similarity  & 98.0 & 97.5 & 97.7 & 98.0 \\
\hline
\end{tabular}
\end{table}
\vspace{-4mm}
{
Table \ref{tab:prompt_results} summarizes the experimental evaluation of the prompts. The simple prompt leads to the lowest performance, with precision at 52.5\%, recall at 40.0\%, F1-score at 45.0\%, and accuracy at 45.0\%, as the model struggles to effectively detect inter-document clones. The structure prompt achieves a higher recall of 80.0\% but is hindered by a high false positive rate, with precision at 60.0\%. This suggests that the model is overly sensitive to structural similarities within the files. The semantic prompt demonstrates moderate performance, with precision at 70.0\%, recall at 60.0\%, F1-score at 64.6\%, and accuracy at 65.0\%. This may be due to the simplified structure of serialized asset files compared to natural language. In contrast, the similarity analysis prompt excels with near-perfect precision at 98\%, recall at 97.5\%, F1-score at 97.7\%, and accuracy at 98.0\%, validating its ability in generating reliable similarity scores for file pairs.}

\begin{table}[htbp]
\caption{\centering{Step 2 of the CoT.}}
\centering
\begin{tabularx}{\textwidth}{|c|c|X|}
\hline
\textbf{Step} & \textbf{Metrics} & \multicolumn{1}{c|}{\textbf{Prompts}} \\
 \hline
\multirow{6}*{2} & \multirow{3}*{\nca, \NCG} & {\em Please analyze the files in the folder and output the number of clone files and clone groups, referencing the cloning information of each file pair from the Output of Step 1. }\\
\cline{2-3}
 & \multirow{3}*{\CI} & {\em Please analyze the files in the folder and output the clone index of the files, referencing the cloning information of each file pair from the Output of Step 1. }\\
\hline
\end{tabularx}
\label{tab:prompts_2}
\end{table}

After calculating the similarity among file pairs, we employ Step 2 of the CoT to compute the \nca, \NCG\ and \CI\ for each project, as detailed in Table \ref{tab:prompts_2}. We set the clone threshold ($\tau'$) to 80\%, following the default recommendation of GPT-4o. The results are subsequently refined through manual inspection and standardization to ensure reliability. Note that the detected serialized files comprise scene files (\verb|.unity|, \verb|.scene|, \verb|.navmesh|), template files (\verb|.prefab|),  material files (\verb|.mat|, \verb|.shader|, \verb|.cginc|), and configuration files (\verb|.json|, \verb|.ini|), but exclude resource files (\verb|.fbx|, \verb|.wav|, \verb|.png|) and \verb|.meta files|. This is because resource files are typically not stored in serialized text format, while \verb|.meta files| are tightly bound to their corresponding assets.

\subsubsection{Metrics for Measuring Code Clone}
\label{sec:metrics}
For source code clone, we focus on code clone at the function granularity. Let $F$ denote the universe of functions in a VR software, and let $\Sim(func_i,func_j)(i \neq j)$ denote the similarity between function $func_i\in F$ and function $func_j\in F$. We define the concepts of {\em clone function} and {\em clone class} below. {It is important to note that the similarity between functions is computed using the NiCad tool, as described in Section \ref{Source_Code_Clone_Detection}.}

\textbf{\emph{Clone function.}} A function $func_i\in F$ is deemed a clone function if there exists at least one other function $func_j\in F (i \neq j)$ such that the similarity between $func_i$ and $func_j$ exceeds a predefined similarity threshold $\tau$. Mathematically, a function $func_i\in F$ is a clone function under the following condition:

\begin{gather}\label{equ_clone_function}
\exists func_j\in F (i \neq j), \Sim(func_i,func_j)> \tau
\end{gather}

\textbf{\emph{Clone class.}} A clone class is defined as a set of two or more functions where the similarity between every pair of functions exceeds a predefined similarity threshold $\tau$. Mathematically, a set $C\subseteq F$ ($|C|\geq 2$) is a clone class under the following condition:
\begin{gather}\label{equ_clone_class}
\forall func_i, func_j\in C (i \neq j), \Sim(func_i,func_j)> \tau
\end{gather}

{ Grounded in the above definitions, we employ four straightforward metrics to evaluate the degree of source code cloning from both absolute and relative perspectives, following the existing studies \cite{roy2008survey, juergens2009code, ain2019systematic, huang2024study}.} The metrics applied to assess the absolute degree of code cloning are as follows:

\begin{itemize}
    \item \textbf{Number of Clone Functions (\ncf):} This metric measures the total number of {\em clone functions} in the source code of a VR software.
    
    \item \textbf{Number of Clone Classes (\ncc):} This metric measures the number of {\em clone classes} in the source code of a VR software.

\end{itemize}
The metrics applied to assess the relative degree of code cloning include:
\begin{itemize}
    \item \textbf{Ratio of Clone Functions (\rcf):} This metric reflects the share of {\em clone functions} in a VR software, which is defined as the ratio of the number of {\em clone functions} (\ncf) to the total number of functions ($|F|$) in the software.

    \item \textbf{Ratio of Clone Classes (\rcc):} This metric reflects the share of {\em clone classes} in a VR software, which is defined as the ratio of the number of {\em clone classes} (\ncc) to the total number of functions ($|F|$) in the software.

\end{itemize}

For serialized asset files clone, we focus on data clone at the file granularity. Let $A$ denote the universe of asset files in a VR software, and let $\Sim'(file_i,file_j)(i \neq j)$ denote the similarity between asset file $file_i\in A$ and asset file $file_j\in A$. We define the concepts of {\em clone file} and {\em clone group} as follows. {It is important to note that the similarity between files is computed using the GPT-4o, as described in Section \ref{Serialization_File_Clone_Detection}.}

\textbf{\emph{Clone file.}} A serialized asset file $file_i\in A$ is deemed a clone file if there exists at least one other asset file $file_j\in A (i \neq j)$ such that the similarity between $file_i$ and $file_j$ exceeds a predefined similarity threshold $\tau'$. Mathematically, a serialized asset file $file_i\in A$ is a clone file under the following condition:
\begin{gather}\label{equ_clone_file}
\exists file_j\in A (i \neq j), \Sim'(file_i,file_j)> \tau'
\end{gather}

\textbf{\emph{Clone group.}} A clone group is defined as a set of two or more serialized asset files where the similarity between every pair of files exceeds a predefined similarity threshold $\tau'$. Mathematically, a set $G\subseteq A$ ($|G|\geq 2$) is a clone group under the following condition:
\begin{gather}\label{equ_clone_group}
\forall file_i, file_j\in A (i \neq j), \Sim'(file_i,file_j)> \tau'
\end{gather}

{Analogous to source code clone metrics, we introduce a set of metrics to evaluate the extent of serialized asset cloning from both absolute and relative perspectives.} The metrics applied to assess the absolute degree of asset cloning are as follows:

\begin{itemize}
    \item \textbf{Number of Clone Assets (\nca):} This metric measures the total number of {\em clone files} in the serialized asset files of a VR software.

\end{itemize}

\begin{itemize}
    \item \textbf{Number of Clone Groups (\NCG):} This metric measures the number of {\em clone groups} in the serialized asset files of a VR software.

\end{itemize}
The metrics applied to assess the relative degree of asset cloning include:
\begin{itemize}
    \item \textbf{Ratio of Clone Assets (\rca):} This metric reflects the share of {\em clone files} in a VR software, which is defined as the ratio of the number of {\em clone files} (\nca) to the total number of asset files ($|A|$) in the software.

\end{itemize}
\begin{itemize}
    \item \textbf{Ratio of Clone Groups (\rcg):} This metric reflects the share of {\em clone groups} in a VR software, which is defined as the ratio of the number of {\em clone groups} (\NCG) to the total number of asset files ($|A|$) in the software.

\end{itemize}

{
To sum up, the proposed metrics (\ncf, \rcf) and (\ncc, \rcc) are based on {\em clone functions} and {\em clone classes}, as defined in Equations (\ref{equ_clone_function}) and (\ref{equ_clone_class}), respectively, in the context of code clones; whereas (\nca, \rca) and (\NCG, \rcg) are based on {\em clone files} and {\em clone groups}, as defined in Equations (\ref{equ_clone_file}) and (\ref{equ_clone_group}), respectively, in the context of file clones.}

To evaluate the overall similarity of asset files in a VR software, we introduce the \textbf{Clone Index (\CI)}, a metric calculated by dividing the sum of pairwise similarities by the total number of asset files, namely

\begin{gather}
\CI= \frac{\sum_{file_i,file_j\in A, i\neq j}^{}\Sim'(file_i,file_j)}{|A|} 
\end{gather}

\subsection{Research Questions Definition}

{The objective of this study is to systematically investigate the phenomenon of code clones in VR software, with the aim of clarifying their characteristics, root causes, and implications. By doing so, we aim to produce actionable insights for researcher, developers, and industry practitioners, ultimately improving the quality, maintainability, and performance of VR software. To achieve this goal, we propose seven research questions (RQs) from three analytical perspectives: (i) general cloning in software, (ii) cloning specific to source code, and (iii) cloning specific to serialized asset files.

\textbf{General cloning in software} (RQ1$\sim$RQ3) provides a comprehensive analysis of the overall cloning landscape in VR software, its intrinsic characteristics, and the measurement methods. 
\begin{itemize}
    \item \textbf{RQ1: Which VR projects are subjected to heavy cloning?} 
   The motivation behind RQ1 is to identify VR projects with significant levels of code and asset cloning, enabling developers to prioritize refactoring efforts and manage technical debt more effectively, providing researchers with insights into why certain projects accumulate more clones, and helping managers allocate resources toward maintaining high-risk systems. To answer this question, we first evaluate how various detection methods influence cloning results, and then assess the extent of cloning across VR projects at both the source code and asset file levels.

    \item \textbf{RQ2: What are the main differences in code cloning between VR and non-VR software?} The motivation behind RQ2 is to understand how the unique structural and interactive features of VR software influence code cloning, since VR's unique demands (e.g., real-time rendering, physics simulation, device I/O) may lead to distinct cloning patterns that differ from traditional software, thereby enabling developers to adopt best practices from non-VR domains where applicable, guiding researchers to design VR-specific clone detection tools if distinctive patterns are observed, and assisting software architects in deciding whether to invest in dedicated infrastructure for VR development. To answer this question, we first analyze the structural distinctions between VR and non-VR software, then assess how these differences contribute to variations in code cloning.

    \item \textbf{RQ3: How to measure and analyze code cloning in VR software development?} The motivation behind RQ3 is to explore and evaluate the suitability of various cloning metrics in the context of VR software, as the unique characteristics of VR systems may require domain-specific measurement approaches, enabling developers to accurately assess code and asset maintainability, and helping researchers and tool builders establish standardized, context-aware evaluation criteria for clone analysis. To answer this question, we offer an in-depth analysis of the physical interpretations and applicability of the proposed metrics for measuring cloning. 
\end{itemize}

\textbf{Cloning specific to source code} (RQ4$\sim$RQ6) explores the factors involved in the introduction of source code cloning.
\begin{itemize}
    \item \textbf{RQ4: Which programming languages dominate VR software development and are more susceptible to code cloning?} The motivation behind RQ4 is to investigate how the choice of programming languages in VR development influences code cloning, since some languages may inherently promote more duplication than others, enabling developers to mitigate language-specific risks through targeted practices and guiding researchers in creating language-aware clone detection methods. To answer this question, we begin by identifying the dominant programming languages in VR development, then investigate how they contribute to code cloning and uncover the root causes behind these patterns.

    \item \textbf{RQ5: How do third-party libraries impact code cloning in VR software?} The motivation behind RQ5 is to examine whether third-party libraries in VR development lead to specific cloning behaviors due to boilerplate or API constraints, helping developers identify dependency-related duplication risks, enabling researchers to distinguish between library-induced and developer-induced cloning patterns for more accurate detection, and informing library authors on improving API design to minimize clone proliferation. To answer this question, we explore the third-party libraries frequently utilized in VR projects and analyze whether their usage leads to code cloning and the specific patterns of cloning that arise.

    \item \textbf{RQ6: How do intra-version and inter-version code cloning evolve across different versions of a VR project?} The motivation behind RQ6 is to examine the temporal dynamics of code cloning in VR projects, as understanding when and why duplication accumulates can help developers schedule refactoring milestones and assist researchers in identifying patterns that inform version-aware maintenance. To answer this question, we investigate the presence of code cloning both within the same software version and between consecutive versions, exploring the degree of cloning and particularly focusing on the differences in cloning across different granularities of version changes. 

\end{itemize}

\textbf{Cloning specific to serialized asset files} (RQ7) examines the influence of asset types on file cloning.
\begin{itemize}
    \item \textbf{RQ7: How do cloning practices vary across different types of asset files in VR software?} The motivation behind RQ7 is to investigate cloning differences among various asset types in VR software, since VR-specific asset types (e.g., 3D models, scenes, shaders) may exhibit different cloning characteristics than traditional software artifacts, helping developers implement asset-specific maintenance strategies, aiding researchers in expanding non-code clone detection in content-rich environments, and assisting asset managers in refining asset control workflows. To answer this question, we analyze the cloning discrepancies of various types of serialized asset files across multiple distinct projects. 

\end{itemize}

}

\newcounter{insightcounter}

\newcommand{\insightbox}[1]{
    \refstepcounter{insightcounter}
    \vspace{0.5em} 
    \textbf{\textsc{Insight \theinsightcounter.}}\hspace{0.5em}{\em #1}\par
    \vspace{0.5em} 
}

\section{Evaluation}
\label{sec:evaluation}

\subsection{Overall Analysis}
{\color{black}
\subsubsection{RQ1: Which VR projects are subjected to heavy cloning?}
To answer this RQ, we utilize NiCad to detect code clones within the VR-345 dataset and present the most cloned projects from the development tools and applications in Table \ref{tab:source_code_different_types_detection}. Specifically, PID 1-10 represent the top ten projects with the highest number of cloned functions in the development tool category, while PID 11-20 are the top ten projects in the VR application category with the most cloned functions. For ease of observation, these projects are arranged in descending order of the total number of functions \(N\) within each category.

We begin by examining the impact of different detection methods on the results, revealing a clear and consistent trend in the $\ncf$ metric (i.e., the number of clone functions): $\ncf$(Type 3-2) \textgreater\ $\ncf$(Type 3-2c) \textgreater\ $\ncf$(Type 3-1), and $\ncf$(Type 2) \textgreater\ $\ncf$(Type 2c) \textgreater\ $\ncf$(Type 1). This suggests that the blind renaming method consistently identifies the most clone fragments, with the consistent renaming method coming in second. This finding exactly aligns with the theoretical expectations outlined in NiCad \cite{cordy2011nicad}, where blind renaming replaces all identifiers with a generic placeholder ``$X_n$'', while consistent renaming assigns sequential identifiers ``$X_n$''($n$ is a sequence number). The rationale is straightforward: if a code fragment is recognized as similar with identifiers renamed as ``$X_n$'',  it will also be recognized as similar when all identifiers are replaced with ``$X$''. Additionally, we observe that $\ncf$(Type 3-x) \textgreater\ $\ncf$(Type x) for x = 1, 2, 2c. This outcome is intuitive since a fragment meeting a 0\% dissimilarity threshold will inherently meet a 30\% threshold. Based on these observations, Type 3-2 emerges as the most effective detection method, with its results encompassing those of other types. In terms of the $\ncc$ metric, there appears to be no clear correlation between the detection types and this metric. As a result, we select Type 3-2 as the standard for code clone detection in the following analyses.

We further observe that the $\ncf$ values for all twenty projects exceed 100, suggesting that code cloning is a common issue in VR project development. Additionally, the highest number of functions ($|F|$) tend to exhibit the largest volume of code clones ($\ncf$). 
Specifically, the top three projects with the most functions are ``BlenderXR", ``gpac", and ``SoundSpace", which also show the highest occurrence of code clones. This observation suggests that, generally, the prevalence of code cloning increases as the scale of a project expands, echoing similar trends found in conventional software development.

\begin{table*}
    \caption{The quantity of clones detected in the source code.}
    \centering
    \resizebox{\textwidth}{!}{
    \begin{tabular}{|c|c|c|c|c|c|c|c|c|c|c|c|c|c|c|c|c|}
    \hline
    \multirow{2}*{\textbf{PID}} & \multirow{2}*{\textbf{Project Name}} & \multirow{2}*{\textbf{Domain}} & \multirow{2}*{\textbf{Language}} & \multicolumn{2}{|c|}{\textbf{Type1}} & \multicolumn{2}{|c|}{\textbf{Type2}} & \multicolumn{2}{|c|}{\textbf{Type2c}} & \multicolumn{2}{|c|}{\textbf{Type3-1}} & \multicolumn{2}{|c|}{\textbf{Type3-2}} & \multicolumn{2}{|c|}{\textbf{Type3-2c}} & \multirow{2}*{\textbf{$|F|$}} \\
    \cline{5-16}
     & & & & \textbf{\ncf} & \textbf{\ncc} & \textbf{\ncf} & \textbf{\ncc} & \textbf{\ncf} & \textbf{\ncc} & \textbf{\ncf} & \textbf{\ncc} & \textbf{\ncf} & \textbf{\ncc} & \textbf{\ncf} & \textbf{\ncc} &  \\
    \hline
    \multicolumn{16}{|c|}{\textbf{Development Tools}} \\
    \hline
    1 & BlenderXR & / & C,Python & 358 & 175 & 1578 & 601 & 1486 & 567 & 1731 & 716 & 4448 & 1226 & 3112 & 889 & 27822 \\
    \hline
    2 & gpac & / & C & 6 & 3 & 1421 & 259 & 1268 & 220 & 1195 & 312 & 2699 & 326 & 2089 & 278 & 10170 \\
    \hline
    3 & The-Seed-Link-Future & / & C\# & 44 & 19 & 231 & 84 & 223 & 85 & 201 & 86 & 462 & 166 & 295 & 112 & 8475 \\
    \hline
    4 & open-brush & / & C\# & 38 & 17 & 132 & 60 & 127 & 58 & 183 & 81 & 409 & 163 & 218 & 89 & 7548 \\
    \hline
    5 & UnityOculusAndroidVRBrowser & / & C\# & 36 & 15 & 156 & 53 & 150 & 55 & 107 & 47 & 255 & 86 & 177 & 65 & 3716 \\
    \hline
    6 & UnityPlugin & / & C\# & 12 & 6 & 53 & 21 & 52 & 21 & 60 & 24 & 113 & 43 & 85 & 33 & 2799 \\
    \hline
    7 & lovr & / & C & 2 & 1 & 84 & 33 & 77 & 30 & 77 & 31 & 232 & 64 & 148 & 53 & 2173 \\  
    \hline
    8 & UnityGameTemplate & / & C\# & 20 & 10 & 60 & 25 & 59 & 25 & 53 & 22 & 142 & 47 & 98 & 35 & 1981 \\
    \hline
    9 & ViveInputUtility-Unity & / & C\# & 20 & 10 & 62 & 27 & 62 & 27 & 69 & 32 & 125 & 49 & 84 & 36 & 1863 \\ 
    \hline
    10 & com.xrtk.core & / & C\# & 3 & 1 & 49 & 18 & 49 & 18 & 42 & 17 & 148 & 51 & 82 & 32 & 1607 \\
    \hline
    \multicolumn{16}{|c|}{\textbf{Applications}} \\
    \hline
    11 & SoundSpace & simulator & C\# &2149&1063&2190&1011&2185&1009&2175&985&2225&965&2201&1001&9117 \\
    \hline
    12 & RhythmAttack-VR & gaming & C\# &37&17&162&60&149&62&118&48&330&119&221&89&6751 \\
    \hline
    13 & Dungeon-VR & gaming & C\# &20&10&100&38&91&35&103&41&254&98&154&61&6470 \\
    \hline
    14 & Situated-Empathy-in-VR & education & C\# &23&11&140&59&136&58&136&59&388&140&262&100&4788 \\
    \hline
    15 & mineRVa & education & C\# &18&9&88&35&82&34&53&24&197&71&140&58&4722 \\
    \hline
    16 & VR-Escape-Room & gaming & C\# &12&6&53&19&44&16&31&13&120&46&77&30&4067 \\
    \hline
    17 & Group6\_ProjectNurture & healthcare & C\# &27&10&148&49&142&51&102&44&248&82&171&62&3980 \\
    \hline
    18 & vrtist & education & C\# &51&17&114&42&112&41&141&48&299&103&187&69&3322 \\
    \hline
    19 & Terminal & gaming & C\# &19&8&96&27&88&29&71&27&160&49&121&41&2929 \\
    \hline
    20 & elite-vr-cockpit & gaming & C\# &0&0&52&18&49&17&30&14&109&39&68&26&2725 \\
    \hline
    \end{tabular}
    }
    \label{tab:source_code_different_types_detection}
\end{table*}

To expand our study on cloning in serialized asset files within VR software, we add 10 more VR applications, covering diverse subcategories like games, education, simulators, and AR, ensuring a comprehensive analysis. Development tools are excluded from the analysis of file cloning because they are typically source-code-centric, with asset files serving configuration and descriptive purposes. In contrast, asset files dominate in application projects, with source code primarily existing as scripts to manipulate these assets. 
Table \ref{tab:serialization_file_clone_detection} provides the clone detection results for the 20 selected VR software applications. 

We observe that the three projects with the highest $\nca$ (i.e., the number of file clones)—``Dungeon-VR", ``Group6\_ProjectNurture" and ``OpendagVR2''—are also the ones with the largest $|A|$ (i.e., the number of asset files), following a pattern similar to source code cloning. {For better clarity, Figure \ref{fig:scatter} visually depicts the relationship between project characteristics and clone metrics, specifically the correlation between the number of asset files ($|A|$) and file clones ($\nca$), with a near-linear relationship observed, suggesting that larger projects typically exhibit more file clones.
}

\begin{figure}
    \centering
    \includegraphics[width=0.7\linewidth]{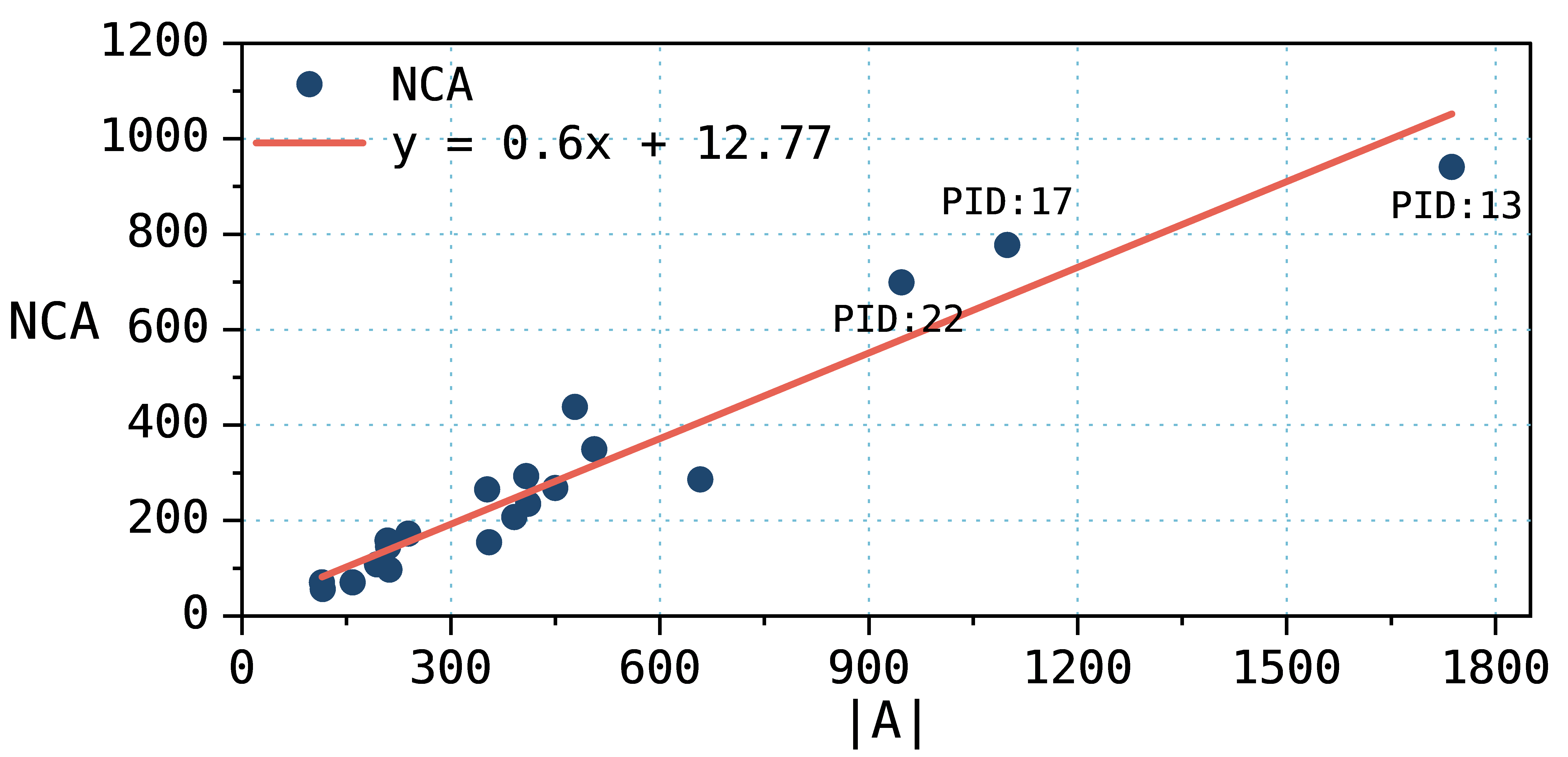}
    \caption{{The relationship between $|A|$ and $\nca$.}}
    \label{fig:scatter}
\end{figure}

\begin{table*}[t]
    \caption{{The quantity of clones detected in serialization files.}}
    \centering
    \resizebox{\textwidth}{!}{
    \begin{tabular}{|c|c|c|c|c|c|c|c|c|c|c|c|}
    \hline
    \multirow{2}*{\textbf{PID}} & \multirow{2}*{\textbf{Project Name}} & \multirow{2}*{\textbf{Domain}} & \multirow{2}*{\textbf{\nca}} & \multicolumn{4}{|c|}{\textbf{\NCG}} & \multirow{2}*{\textbf{$|A|$}} & \multirow{2}*{\textbf{\rca}}& \multirow{2}*{\textbf{\rcg}}& \multirow{2}*{\textbf{\CI}} \\
    \cline{5-8}
    & & & & \textbf{2} & \textbf{[3, 10]} & \textbf{\textgreater\ 10} & \textbf{sum} & & & &\\
    \hline
    \hline
    12 & RhythmAttack-VR & gaming & 286 & 20 & 19 & 5 & 44 & 658 & 43.47\% & 6.69\% & 8.17\% \\ 
    \hline
    13 & Dungeon-VR & gaming & 941 & 31 & 20 & 15 & 66 & 1737 & 54.17\% & 3.80\% & 16.32\% \\ 
    \hline
    16 & VR-Escape-Room & gaming & 154 & 10 & 6 & 4 & 20 & 355 & 43.38\% & 5.63\% & 9.69\% \\ 
    \hline
    19 & Terminal & gaming & 268 & 11 & 13 & 4 & 28 & 450 & 59.56\% & 6.22\% & 23.18\% \\ 
    \hline
    20 & elite-vr-cockpit & gaming & 97 & 5 & 8 & 1 & 14 & 212 & 45.75\% & 6.60\% & 12.35\% \\ 
    \hline
    21 & VRHamsterBall & gaming & 108 & 5 & 10 & 1 & 16 & 194 & 55.67\% & 8.25\% & 16.37\% \\ 
    \hline
    23 & epicslash & gaming & 145 & 7 & 8 & 6 & 21 & 210 & 69.05\% & 10.00\% & 17.46\% \\ 
    \hline
    26 & AirAttack & gaming & 265 & 8 & 9 & 5 & 22 & 352 & 75.28\% & 6.25\% & 27.28\% \\ 
    \hline
    27 & GolfVR & gaming & 56 & 4 & 3 & 2 & 9 & 116 & 48.28\% & 7.76\% & 13.84\% \\ 
    \hline
    28 & XR-Keyboard & gaming & 70 & 5 & 2 & 2 & 9 & 159 & 44.03\% & 5.66\% & 15.18\% \\ 
    \hline
    29 & HorrorGame & gaming & 349 & 12 & 27 & 7 & 46 & 506 & 68.97\% & 9.09\% & 13.93\% \\ 
    \hline
    30 & Pokemon-Themed-Kiosk-VR & gaming & 172 & 10 & 4 & 5 & 19 & 239 & 71.97\% & 7.95\% & 22.96\% \\ 
    \hline
    \hline
    14 & Situated-Empathy-in-VR & education & 158 & 1 & 3 & 2 & 6 & 209 & 75.60\% & 2.87\% & 57.42\% \\ 
    \hline
    15 & mineRVa & education & 235 & 8 & 8 & 3 & 19 & 411 & 57.18\% & 4.62\% & 18.67\% \\ 
    \hline
    18 & vrtist & education & 293 & 21 & 17 & 5 & 43 & 408 & 71.81\% & 10.54\% & 14.89\% \\ 
    \hline
    22 & OpendagVR2 & education & 699 & 21 & 26 & 10 & 57 & 947 & 73.81\% & 6.02\% & 24.05\% \\ 
    \hline
    25 & Procrastination-VR & education & 438 & 5 & 5 & 2 & 12 & 478 & 91.63\% & 2.51\% & 69.29\% \\ 
    \hline
    \hline
    17 & Group6\_ProjectNurture & healthcare & 777 & 24 & 15 & 12 & 51 & 1099 & 70.70\% & 4.64\% & 23.28\% \\ 
    \hline
    \hline
    11 & SoundSpace & simulator & 207 & 10 & 15 & 3 & 28 & 391 & 52.94\% & 7.16\% & 11.24\% \\ 
    \hline
    24 & CityMatrixAR & simulator & 70 & 3 & 5 & 2 & 10 & 115 & 60.87\% & 8.70\% & 25.13\% \\ 
    \hline
    \end{tabular}}
    \label{tab:serialization_file_clone_detection}
\end{table*}

\insightbox{Both source code cloning and serialized file cloning in VR software show a consistent pattern, with larger projects generally exhibiting more clones.}

{When examining the domains of the top 20 most clone-prone software systems, gaming and education emerge as the most impacted. According to Table \ref{tab:source_code_different_types_detection} and Table \ref{tab:serialization_file_clone_detection}, code cloning occurs in 5 gaming and 3 education systems, while file cloning is present in 12 gaming and 5 education systems. This recurring pattern highlights the need for special attention to cloning control in these areas, especially in gaming.

\insightbox{Code and file cloning issues are most prevalent in gaming and education software, with gaming being the most significantly impacted, highlighting the need for targeted attention in this domain.}

}

\subsubsection{RQ2: What are the main differences in code cloning between VR and non-VR software? }
{To answer this RQ, we start by reviewing prior work on code cloning in traditional software. Among the most relevant studies, {Kamiya et al. \cite{kamiya2002ccfinder} reported a cloned function ratio of 12\% to 16\% in medium and large software}. Kapser et al. \cite{kapser2006cloning} found a similar ratio, close to 20\%, in C-based projects. Roy et al. \cite{roy2008survey} observed that Java and C++ projects exhibited a clone function ratio ranging from 10\% to 15\%. These findings suggest that code cloning is a widespread and consistent issue across traditional software development, typically affecting 10\% to 20\% of functions.

To investigate code cloning in VR software, we propose exploring the relationship between the ratio of clone functions ($\rcf$) and the quantity of asset files ($|A|$), as the use of such assets constitutes a key difference between VR and traditional software. As shown in Table \ref{tab:NCF_and_|A|}, we observe that the development tools Project-1 (``BlenderXR") and Project-2 (``gpac"), along with the VR applications Project-11 (``SoundSpace") and Project-24 (``CityMatrixAR") exhibit significantly higher levels of source code cloning than other projects, with $\rcf$ values ranging from 16\% to 27\%, whereas most of the remaining 26 projects fall below 10\%.

We further examine the four outliers with unusually high levels of code cloning. Project-1 and Project-2 lack any serialized asset files. Project-11 and Project-24, though containing some serialized files, have relatively few when compared to the overall volume of source code. Specifically, Project-24, an AR application, constructs models from real-world inputs rather than serialized data. Similarly, Project-11, which focuses on acoustic simulation and visualization, does not require extensive use of serialized assets. A shared trait among these projects is the limited presence of serialized files, which aligns them more closely with traditional software patterns. As a result, they exhibit cloning behavior similar to that observed in traditional software, with clone function ratios exceeding 10\%.

By comparison, VR software projects that utilize extensive serialized asset files demonstrate markedly reduced code cloning, as evidenced by the remaining 26 projects' average clone function ratio of 5.02\%. This may result from the limited capability of clone detection tools (e.g., NiCad) to recognize duplicates in serialized files, which differ structurally from standard source code. It is important to note that Project-7 (``lovr"), although written in C, also contains many model class files (\verb|/modules|), which contributes to its relatively lower levels of code cloning. This observation does not contradict our conclusions. In summary, asset files significantly influence code cloning patterns, presenting unique detection challenges that current clone analysis tools must address for effective VR software maintenance.

\begin{table*}[t]
    \caption{{The correlation between the ratio of clone functions ($\rcf$) and the number of asset files ($|A|$) across VR projects.}}
    \centering
    \resizebox{\textwidth}{!}{
    \begin{tabular}{|c|c|c|c|c|c|c|c|c|c|c|}
    \hline
    \textbf{PID} & \textbf{1} & \textbf{2} & \textbf{3} & \textbf{4} & \textbf{5} & \textbf{6} & \textbf{7} & \textbf{8} & \textbf{9} & \textbf{10} \\
    \hline
    \textbf{\rcf} &16.00\% & 26.54\% & 5.45\% & 5.42\% & 6.86\% & 4.04\% & 10.68\% & 7.17\% & 6.71\% & 9.21\% \\
    \hline
    \textbf{$|A|$} &0&0&303&1037&321&222&0&63&93&8 \\
    \hline
    \hline
    \textbf{PID} & \textbf{11} & \textbf{12} & \textbf{13} & \textbf{14} & \textbf{15} & \textbf{16} & \textbf{17} & \textbf{18} & \textbf{19} & \textbf{20} \\
    \hline
    \textbf{\rcf} &24.41\% & 4.89\% & 3.93\% & 8.10\% & 4.17\% & 2.95\% & 6.23\% & 9.00\% & 5.46\% & 4.00\% \\
    \hline
    \textbf{$|A|$} &391&658&1737&209&411&355&1099&408&450&212 \\
    \hline
    \hline
    \textbf{PID} & \textbf{21} & \textbf{22} & \textbf{23} & \textbf{24} & \textbf{25} & \textbf{26} & \textbf{27} & \textbf{28} & \textbf{29} & \textbf{30} \\
    \hline
    \textbf{\rcf} &3.65\% & 2.52\% & 4.11\% & 18.20\% & 7.45\% & 9.88\% & 7.11\% & 2.65\% & 3.87\% & 0.79\% \\
    \hline
    \textbf{$|A|$} &194&947&210&115&478&352&116&159&506&239 \\
    \hline
    \end{tabular}}
    \label{tab:NCF_and_|A|}
\end{table*}

\insightbox{Unlike traditional software, most VR projects exhibit significantly lower source code cloning due to their heavy reliance on serialized asset files.
}
}

{\color{black}
\subsubsection{RQ3: {How to measure and analyze code cloning in VR software development?}
}

To answer this RQ, we sequentially analyze the applicability of the metrics defined in Section \ref{sec:metrics}. 

{\ncf} and {\ncc}. The physical interpretation of $\ncf$ is relatively intuitive; higher values directly indicate a greater cloning volume. However, the implication of $\ncc$ is less straightforward. Using Type3-2 in Table \ref{tab:source_code_different_types_detection} as an example, we observe a notable difference between {\ncf} and {\ncc}. For instance, comparing Project-2 to Project-1, the $\ncf$ ratio is approximately $2699/4448\approx60.7\%$, whereas the $\ncc$ ratio is considerably lower at $326/1226\approx26.6\%$. In contrast, when comparing Project-4 to Project-3, the 
$\ncf$ ratio is $409/462\approx88.5\%$, but the $\ncc$ ratio is even higher at $163/166\approx98.2\%$. To uncover the underlying cause of these differences, we analyze the correlation between the number of clone classes and their respective sizes. As shown in Table \ref{tab:NCC_distribution}, Project-2 exhibits a significantly higher proportion of large-size (\textgreater100) and mid-size ([11, 100]) classes compared to Project-1, leading to a notable reduction in \ncc. However, Project-3 and Project-4 have similar clone class distributions across different intervals, which results in a close $\ncc$. The observations imply that $\ncc$ acts as a valuable indicator of clone class size, with lower values typically reflecting the presence of larger clone classes.

\begin{table}[t]
    \caption{The distribution of the sizes of clone classes.}
    \centering
    \begin{tabular}{|c|c|c|c|c|c|} 
    \hline 
    \diagbox{\textbf{Project Name}}{\textbf{NCC Interval}}& \textbf{2} & \textbf{[3,10]} & \textbf{[11, 100]} & \textgreater\ \textbf{100} & \textbf{Sum} \\ 
    \hline
    BlenderXR (PID:1) & 898(73.2\%) & 304(24.8\%) & 23(1.9\%) & 1(0.1\%) & 1226 \\
    \hline
    gpac (PID:2) & 207(63.5\%) & 100(30.7\%) & 13(4.0\%) & 6(1.8\%) & 326\\
    \hline
    The-Seed-Link-Future (PID:3) & 122(73.5\%) & 39(23.5\%) & 5(3.0\%) & 0 & 166\\
    \hline
    open-brush (PID:4) & 123(75.5\%) & 37(22.7\%) & 3(1.8\%) & 0 & 163\\
    \hline 
    \end{tabular}
    \label{tab:NCC_distribution}
\end{table}

\begin{figure}[t]
    \centering
    \includegraphics[width=\linewidth]{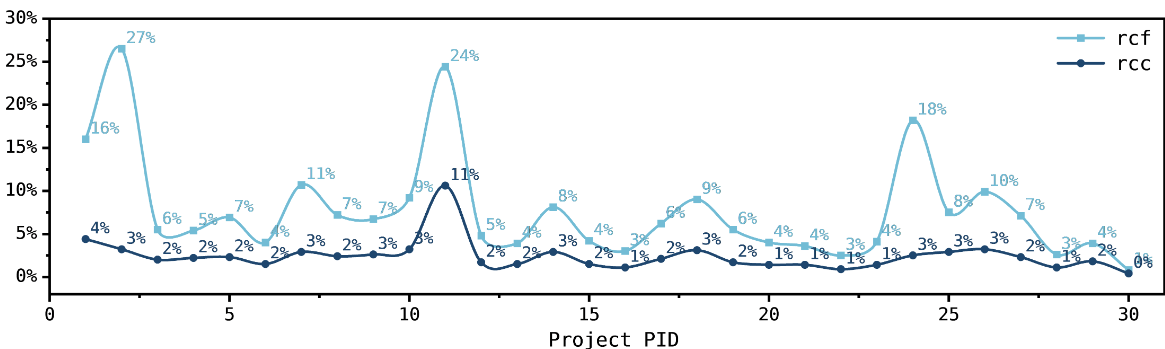}
    \caption{The changes of $\rcf$ and $\rcc$ across diverse projects.}
    \label{fig:rcf_rcc}
\end{figure}

{\rcf} and {\rcc.} To explore the implications of \rcf (i.e., the proportion of clone functions, representing the density of function-level code clones) and \rcc (i.e., the proportion of unique clone classes, representing the diversity of code clone patterns), we analyze their variations across different projects, as illustrated in Figure \ref{fig:rcf_rcc}. The results reveal that $\rcf$ exhibits significant fluctuations, while $\rcc$ remains relatively consistent across projects except for Project-11. The anomaly in the project ``SoundSpace" arises from two highly similar directories, ``SoundSpace" and ``rv\_SoundSpace", with 902 of 965 clone classes being pairwise clones. These predominantly size-2 clones lead to an unusually high $\rcc$. Additionally, no clear correlation is observed between $\rcf$ and $\rcc$, suggesting that the density of function-level code clones and the diversity of clone patterns are largely independent of each other. This decoupling highlights the complexity of code clone characteristics and underscores the need for separate consideration of density and diversity in clone analysis. 

\begin{figure}
    \centering
    \includegraphics[width=\linewidth]{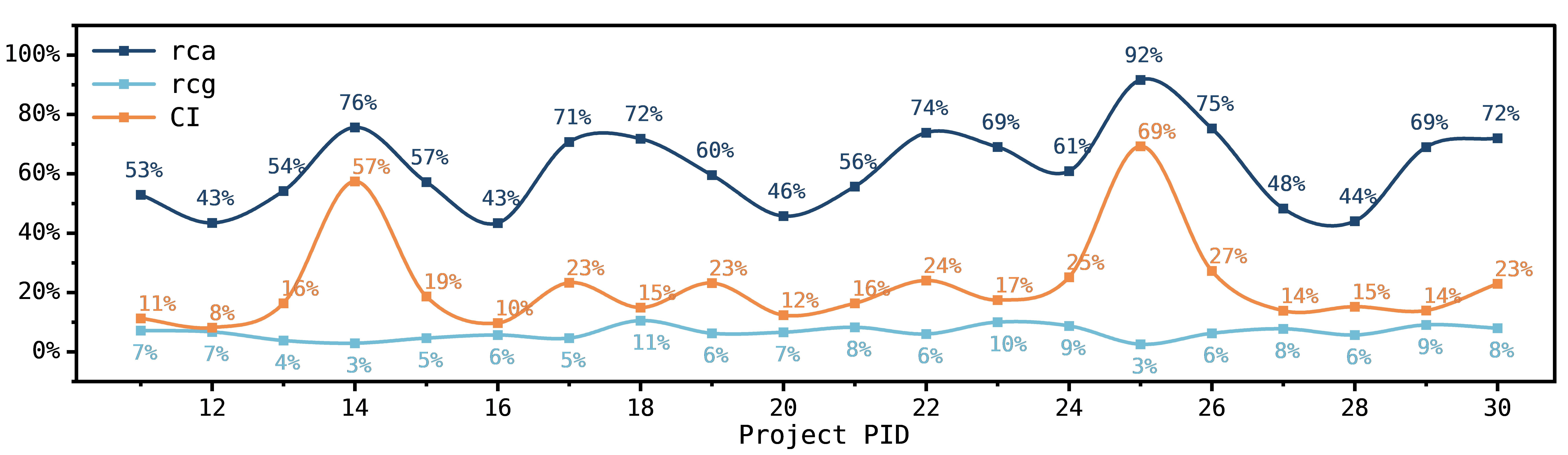}
    \caption{The changes of {\rca}, {\rcg} and {\CI} across diverse projects.}
    \label{fig:enter-label}
\end{figure}

{\nca} and {\NCG.} The meanings of $\nca$ and $\NCG$ are analogous to $\ncf$ and $\ncc$ in source code cloning. A higher $\nca$ indicates a greater quantity of cloned assets within the project, while the impact of $\NCG$ is reflected in the concentration of clones. Given a fixed $\nca$ value, a lower $\NCG$ suggests that the number of members in individual clone groups is higher, indicating more concentrated cloning within the project. 

{\rca}, {\rcg} and {\CI.} To explore the meanings of these three metrics, we plot their respective values for the 20 selected projects, as shown in Figure \ref{fig:enter-label}. The results indicate that $\rca$ and $\CI$ exhibit a certain degree of correlation, while $\rcg$ remains relatively stable across the 20 projects. We observe that Project-14 (``Situated-Empathy-in-VR") and Project-25 (``Procrastination-VR") show exceptionally high $\CI$, standing out significantly from other projects. Upon further investigation, we find that these two projects have the highest $\rca$ values (75.60\% and 91.63\%) among all projects, while their $\rcg$ values are the lowest (2.87\% and 2.51\%), meaning that they possess the most cloned files but the fewest clone groups. This corresponds to the case where a few files are heavily cloned, leading to the emergence of some exceptionally large clone groups. Thus, we can deduce that $\CI$ is jointly influenced by $\rca$ and $\rcg$, where $\rca$ shows a positive correlation and $\rcg$ a negative correlation.

\insightbox{The proposed metrics measure cloning from distinct perspectives: the number or ratio of clone functions (files) measures the quantity of clones, while the number or ratio of clone classes (groups) evaluates clone concentration. By combining these metrics, a comprehensive clone evaluation can be achieved.}
}

\subsection{Source Code Cloning Analysis}

\subsubsection{RQ4: Which programming languages dominate VR software development and are more susceptible to code cloning?}
To answer this RQ, we visualize the programming languages used by these software projects in Figure \ref{fig:language_distribution}. The left pie chart shows the programming language distribution for all VR-345 projects, while the right highlights filtered high-star projects with star \textgreater= 200. 
In this context, the programming language mentioned for a project refers to the primary language used within the project. 

\begin{figure}[t]
    \centering
    \begin{minipage}{0.39\columnwidth}
        \centering
         \includegraphics[width=\columnwidth]{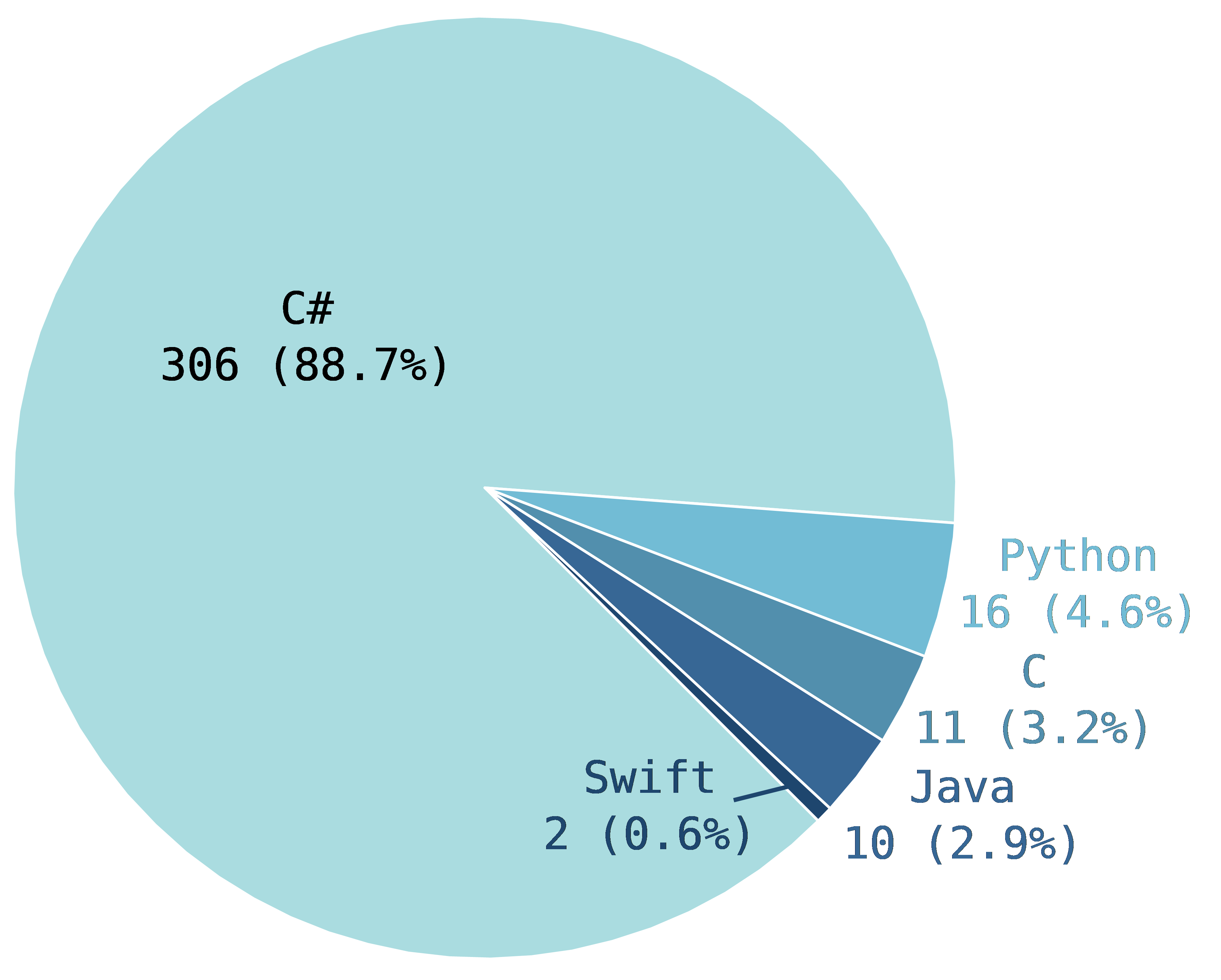}
        \subcaption{All VR-345 projects.}
        \label{fig:language_all}
    \end{minipage}
    \hspace{0.07\columnwidth}
    \begin{minipage}{0.36\columnwidth}
        \centering
        \includegraphics[width=0.89\columnwidth]{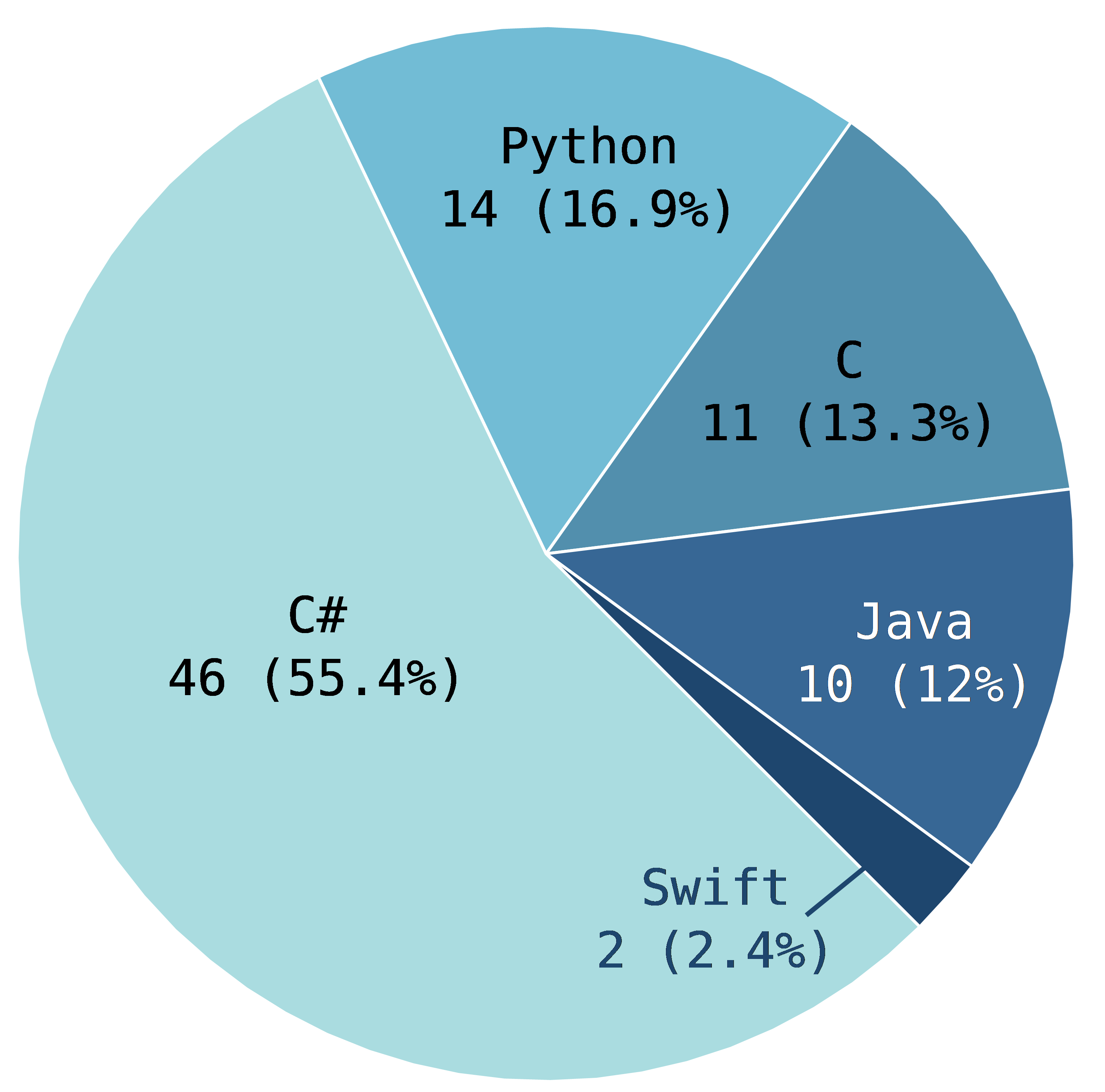}
        \subcaption{83 high-star projects.}
        \label{fig:language_star}
    \end{minipage}
    \caption{Distribution of programming languages in VR projects.}
    \label{fig:language_distribution}
\end{figure}

We observe that C\# dominates general VR projects, while it accounts for slightly over half in high-star projects. This is due to the fact that high-star projects not only feature engine-based applications but also include a substantial number of development tool projects, which are typically developed in various programming languages and serve as support for other projects. The advantages of C\# language are most likely attributed to its close integration with the Unity game engine, which has emerged as one of the leading tools in VR development due to its ease of use and cross-platform
features. 
Given that C\# serves as the primary programming language for Unity, it is unsurprising that it has become the most widely adopted language for VR projects. Other languages employed in VR development include Python, C, Java, and Swift. In contrast, languages such as PHP, Ruby, WSDL, and ATL are absent. This is reasonable because web-oriented languages like PHP, Ruby, and WSDL are not well-suited for graphics-intensive applications due to limitations in execution efficiency and performance, while ATL is primarily designed for Windows-specific programming. It is also worth noting that JavaScript plays a significant role in VR project development. However, its absence in this analysis stems from a limitation in NiCad, which currently does not support clone detection for JavaScript.

To further examine which programming languages are more prone to code cloning, we identify projects with 10 or more clone classes (\ncc \textgreater= 10) in terms of Clone Type 3-2 within high-star VR projects and categorize these projects by their primary programming languages. Table \ref{tab:language_ncc} presents the distribution of projects across various languages at different cloning levels. The results reveal that C\# accounts for the highest number of projects (18) with over 10 clone classes, followed by C with 5 projects. This finding aligns with the prominent role of C\# as the most commonly used language in VR software development. However, when considering the cloning ratio, C ($5/11\approx45.5\%$) surpasses C\# ($18/46\approx39.1\%$), indicating that C may have a higher likelihood of triggering cloning issues. {This may stem from the fact that C, being process-oriented, does not support the code reuse mechanisms inherent to object-oriented languages, so developers might inadvertently copy blocks of code instead of abstracting them.} 

{
\insightbox{While C\# is the most commonly used programming language in open-source VR projects, C-based projects exhibit a higher proportion of code cloning.}}

\begin{table}[t]
    \caption{The distribution of projects across different cloning intervals with regard to $\ncc$ using various languages.}
    \centering
    \begin{tabular}{|c|c|c|c|c|c|} 
    \hline 
    \diagbox{\textbf{Language}}{\textbf{NCC Interval}}& \textbf{[10, 30]} & \textbf{[31, 50]} & \textbf{[51, 100]} & \textgreater\ \textbf{100} & \textbf{Sum}\\
    \hline
    Java & 1 & 0 & 0 & 0 & 1 (out of 10)\\
    \hline
    Python & 3 & 0 & 0 & 1 & 4 (out of 14)\\
    \hline
    C & 1 & 1 & 1 & 2 & 5 (out of 11)\\
    \hline
    C\# & 7 & 7 & 2 & 2 & 18 (out of 46)\\
    \hline 
    \end{tabular}
    \captionsetup[table]{ labelsep=newline, singlelinecheck=false}
    \label{tab:language_ncc}
\end{table}

\subsubsection{RQ5: How do third-party libraries impact code cloning in VR software?}

\begin{table*}[t]
\caption{The list of analyzed third-party libraries.}
\centering
\resizebox{\textwidth}{!}{
\begin{tabular}{|c|c|c|c|c|}
\hline
\textbf{Library} & \textbf{Type} & \textbf{Key Files} & \textbf{Exist} & \textbf{Clone} \\
\hline
SteamVR & VR Hardware SDK & \texttt{Assets/SteamVR/Input/SteamVR\_Input.cs} & \checkmark & \checkmark \\
\hline
Oculus & VR Platform SDK & \texttt{Assets/Oculus/VR/Scripts/OVRManager.cs} & \checkmark & \checkmark \\
\hline
OpenVR & VR Runtime & \texttt{Assets/Plugins/openvr/api.cs} & \checkmark & × \\
\hline
Vive & Hardware Interaction SDK & \texttt{Assets/SteamVR/InteractionSystem/Core/Interfaces/Hand.cs} & \checkmark & × \\
\hline
Google VR & Mobile VR SDK & \texttt{Assets/GoogleVR/Legacy/Scripts/GvrViewer.cs} & \checkmark & × \\
\hline
VRTK & VR Framework & \texttt{Assets/VRTK/SDK/Base/Scripts/VRTK\_SDKManager.cs} & \checkmark & \checkmark \\
\hline
Unity XR & Cross-Platform SDK & \texttt{Assets/XR/Input/XRController.cs} & × &  \\
\hline
OpenXR & Cross-Vendor Runtime & \texttt{Assets/XR/OpenXR/Settings/OpenXRSettings.asset} & × &  \\
\hline
OSVR & VR Platform & \texttt{Assets/OSVR/ClientKit/OSVR\_ClientContext.cs} & × &  \\
\hline
Pico SDK & VR SDK & \texttt{Assets/PicoMobileSDK/Scripts/PicoVR\_Manager.cs} & × &  \\
\hline
Wave SDK & VR Platform & \texttt{Assets/WaveVR/Scripts/WaveVR\_Reticle.cs} & × &  \\
\hline
Varjo Base & Enterprise VR SDK & \texttt{Assets/Varjo/Scripts/VarjoManager.cs} & × &  \\
\hline
Ultraleap & Gesture SDK & \texttt{Assets/Ultraleap/Hands/LeapHandController.cs} & × &  \\
\hline
WebXR & Browser-Based VR & \texttt{Assets/WebXR/Plugins/WebXRInterface.cs} & × &  \\
\hline
\end{tabular}}
\label{tab:3rd_party_file}
\end{table*}

\begin{table*}[t]
    \caption{The evaluation results of third-party libraries in the projects.}
    \centering
    \resizebox{\textwidth}{!}{
    \begin{tabular}{|c|>{\centering\arraybackslash}p{3.6cm}|c|c|c|c|c|c|c|}
            \hline
            \multirow{2}*{\textbf{PID}} & \multicolumn{3}{|c|}{\textbf{Project}} & \multicolumn{5}{|c|}{\textbf{Library}} \\
            \cline{2-9}
               & \textbf{Name} & \textbf{NCF} & \textbf{NCC} & \textbf{Name} & \textbf{NCF} & \textbf{Per} & \textbf{NCC} & \textbf{Per} \\
            \hline
            11 & SoundSpace & 2225 & 965 & SteamVR & 70 & 3.17\% & 26 & 2.69\% \\
            \hline
            \multirow{3}*{12} & \multirow{3}*{RhythmAttack-VR} & \multirow{3}*{330} & \multirow{3}*{119} & SteamVR & 29 & \multirow{3}*{73.64\%} & 8 & \multirow{3}*{68.07\%} \\
            \cline{5-6} \cline{8-8}
               &&&  & Oculus & 54 && 13 & \\
            \cline{5-6} \cline{8-8}
               &&&  & VRTK & 160 && 60 & \\
            \hline
            \multirow{2}*{13} & \multirow{2}*{Dungeon-VR} & \multirow{2}*{254} & \multirow{2}*{98} & SteamVR & 29 & \multirow{2}*{57.87\%} & 8 & \multirow{2}*{42.86\%} \\
            \cline{5-6} \cline{8-8}
               &&&  & VRTK & 81 && 34 & \\
            \hline
            \multirow{2}*{15} & \multirow{2}*{mineRVa} & \multirow{2}*{197} & \multirow{2}*{71} & SteamVR & 33 & \multirow{2}*{97.46\%} & 8 & \multirow{2}*{94.37\%} \\
            \cline{5-6} \cline{8-8}
               &&&  & VRTK & 158 && 59 & \\
            \hline
            \multirow{2}*{16} & \multirow{2}*{VR-Escape-Room} & \multirow{2}*{120} & \multirow{2}*{46} & SteamVR & 29 & \multirow{2}*{91.67\%} & 8 & \multirow{2}*{91.30\%} \\
            \cline{5-6} \cline{8-8}
               &&&  & VRTK & 81 && 34 & \\
            \hline
            17 & Group6\_ProjectNurture & 248 & 82 & Oculus & 222 & 89.52\% & 69 & 84.15\% \\
            \hline
            \multirow{2}*{19} & \multirow{2}*{Terminal} & \multirow{2}*{160} & \multirow{2}*{49} & SteamVR & 22 & \multirow{2}*{56.88\%} & 6 & \multirow{2}*{46.94\%} \\
            \cline{5-6} \cline{8-8}
               &&&  & Oculus & 69 && 17 & \\
            \hline
            20 & elite-vr-cockpit & 109 & 39 & SteamVR & 83 & 76.15\% & 28 & 71.79\% \\
            \hline
            21 & VRHamsterBall & 74 & 28 & SteamVR & 66 & 89.19\% & 24 & 85.71\% \\
            \hline
            22 & OpendagVR2 & 45 & 16 & SteamVR & 29 & 64.44\% & 8 & 50.00\% \\
            \hline
            23 & epicslash & 54 & 19 & SteamVR & 26 & 48.15\% & 7 & 36.84\% \\
            \hline
            26 & AirAttack & 94 & 30 & Oculus & 54 & 57.45\% & 13 & 43.33\% \\
            \hline
        \end{tabular}}
    \label{tab:3rd_party_evaluation}
\end{table*}

{{To answer this RQ, we examine 20 application projects to identify third-party libraries with code clones. We initially identify 14 widely used libraries, 6 of which are found in our dataset, with 3—SteamVR, Oculus, and VRTK—showing cloning issues. The specific procedure is as follows: (i) we identify 14 widely used third-party libraries in VR development for study, as shown in Table \ref{tab:3rd_party_file}; (ii) we verify whether each of these 14 libraries is integrated in any of the 20 projects by checking for the existence of their distinctive key files, resulting in 6 libraries being present; (iii) we then detect whether the remaining libraries have code clones, identifying 3 libraries—SteamVR, Oculus, and VRTK—with cloning issues; and (iv) we review the cloning results to ensure all libraries with clones are properly examined.}

Upon reviewing our results, we find that SteamVR appears in 10 projects, while Oculus and VRTK each appear in 4 projects. However, when it comes to introducing source code clones, SteamVR has the lowest contribution, Oculus ranks in the middle, and VRTK contributes the most. This is reasonable because VRTK, as an open-source toolkit, supports numerous VR devices and platforms, providing high-level abstractions that simplify development and lead to significant code redundancy. On the other hand, SteamVR and Oculus are development kits designed for specific hardware, prioritizing compatibility and low-level optimization, which results in less cross-platform or high-level redundant code. }

These findings confirm the presence of code clones caused by third-party libraries in VR software. Such clones are not introduced by the developers but are intrinsic to the third-party libraries themselves. For instance, in the three third-party libraries identified in our analysis,  different projects display the same cloning results, such as ($\ncf$:29, $\ncc$:8) of SteamVR in Project-12, 13, 16, 22, ($\ncf$:54, $\ncc$:13) of Oculus in Project-12, 26, and ($\ncf$:81, $\ncc$:34) of VRTK in Project-13, 16. 
These clones are triggered by the most commonly used code in these libraries, with additional clones being introduced by project-specific features. 
Although developers cannot modify clones in third-party tools, they can minimize them by using dependency injection, interface-based techniques, or opting for lightweight libraries when cross-platform support is unnecessary.

\insightbox{Third-party libraries introduced in VR applications often contribute to source code cloning issues, sometimes constituting the majority of the cloning results. The more functionalities the third-party libraries provide, the higher the clone evaluation metrics tend to be.}

{\color{black}
\subsubsection{RQ6: How do intra-version and inter-version code cloning evolve across different versions of a VR project?}

To answer this RQ, we begin by examining whether code clones exist between different versions of the same project. To achieve this, we select two projects with a high number of versions and a wide range between versions for analysis. Specifically, We execute code clone detection on six sequential versions of each project, with the detection carried out between each pair of adjacent versions, yielding five data sets per project. The experimental results are illustrated in Figure \ref{fig:inter_version}.

\begin{figure}
    \centering
    \begin{minipage}{0.47\columnwidth}
        \centering
        \includegraphics[width=\columnwidth]{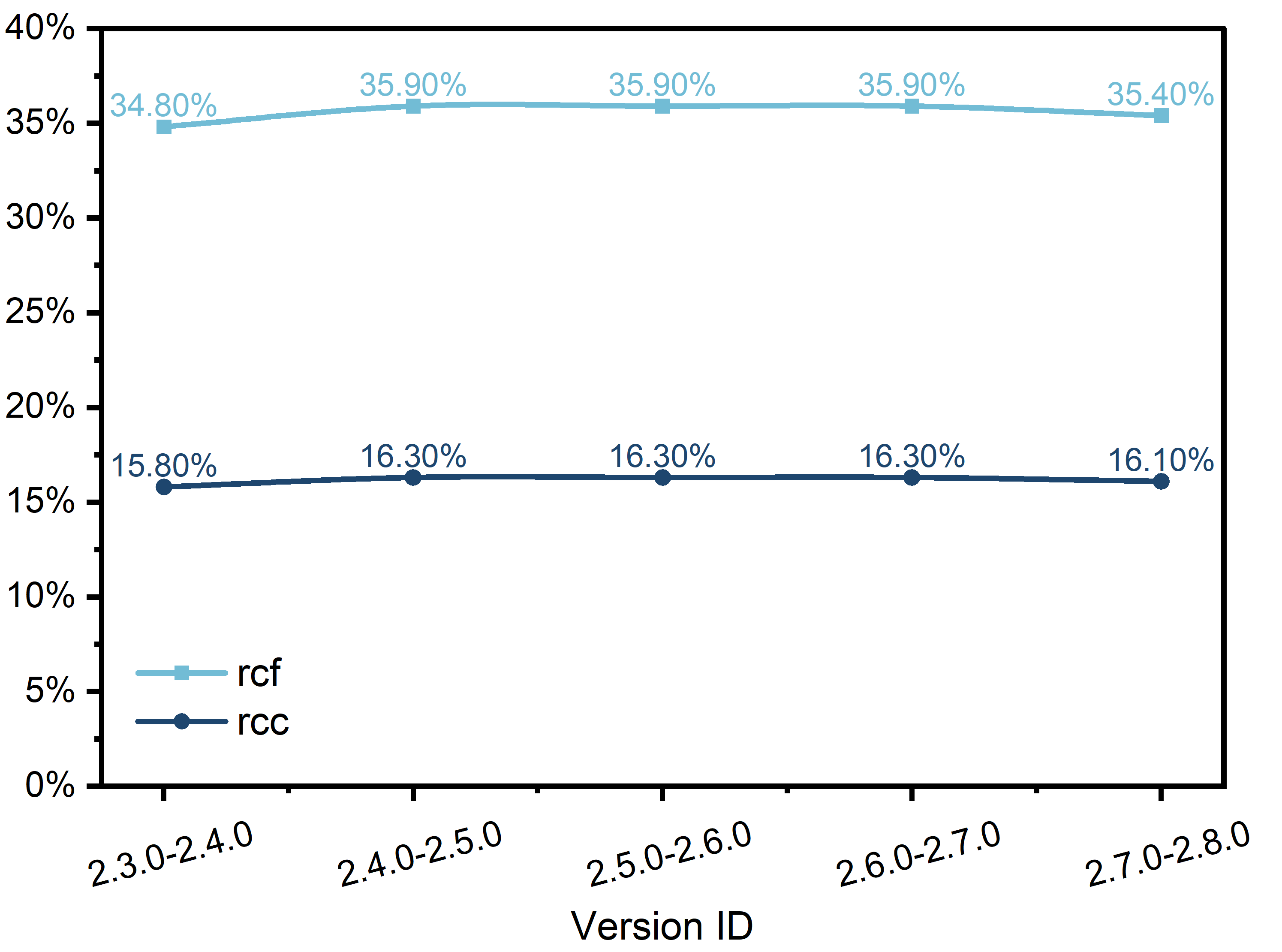}
        \subcaption{open-brush}
        \label{fig:inter_open-brush}
    \end{minipage}
    \hspace{0.01\columnwidth}
    \begin{minipage}{0.47\columnwidth}
        \centering
        \includegraphics[width=\columnwidth]{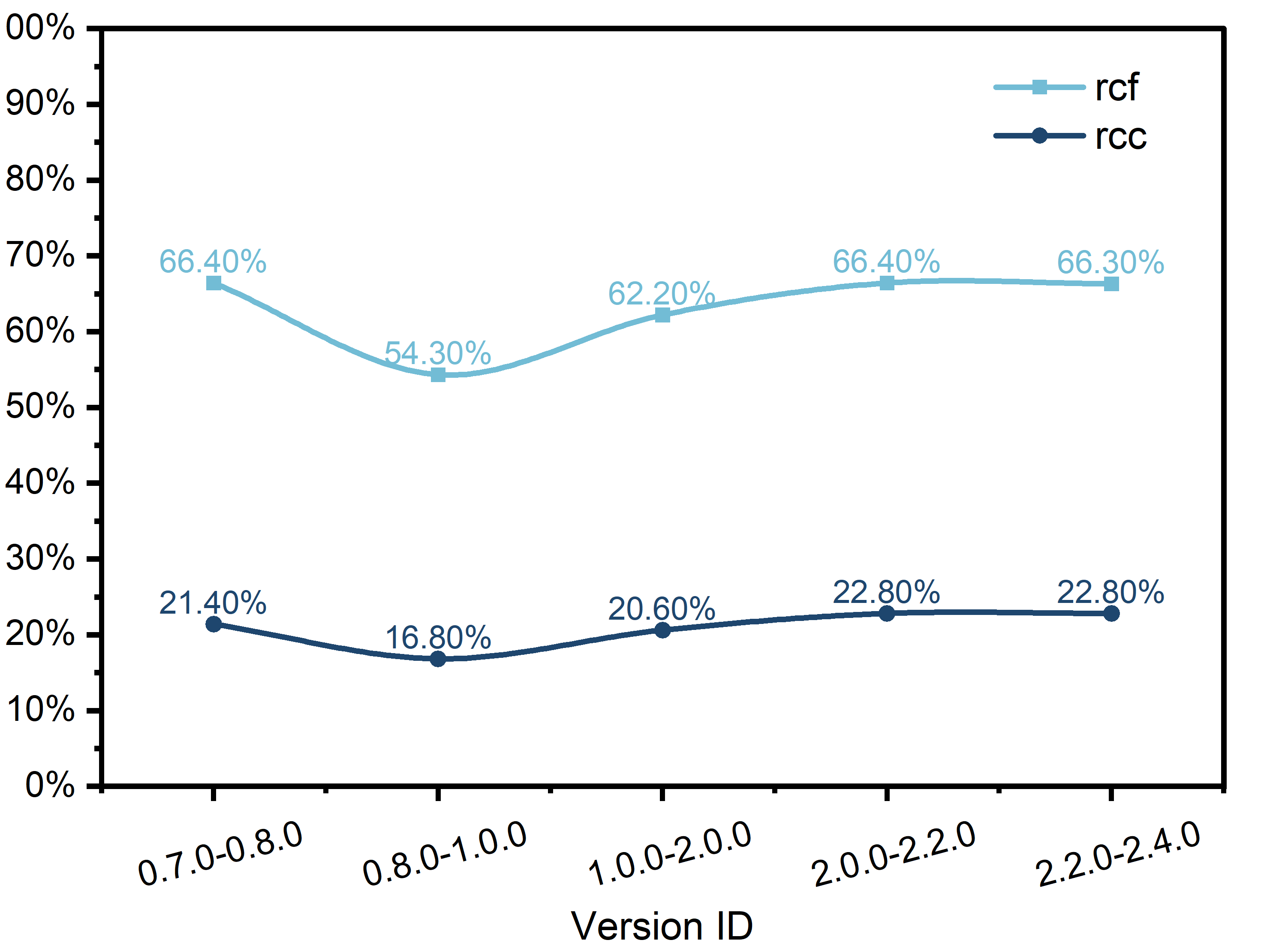}
        \subcaption{gpac}
        \label{fig:gpac}
    \end{minipage}
    \caption{Detection results of inter-version code clones on five adjacent version pairs of two projects.}
    \label{fig:inter_version}
\end{figure}

\begin{figure}[t]
    \centering
    \begin{minipage}{0.47\columnwidth}
        \centering
        \includegraphics[width=\columnwidth]{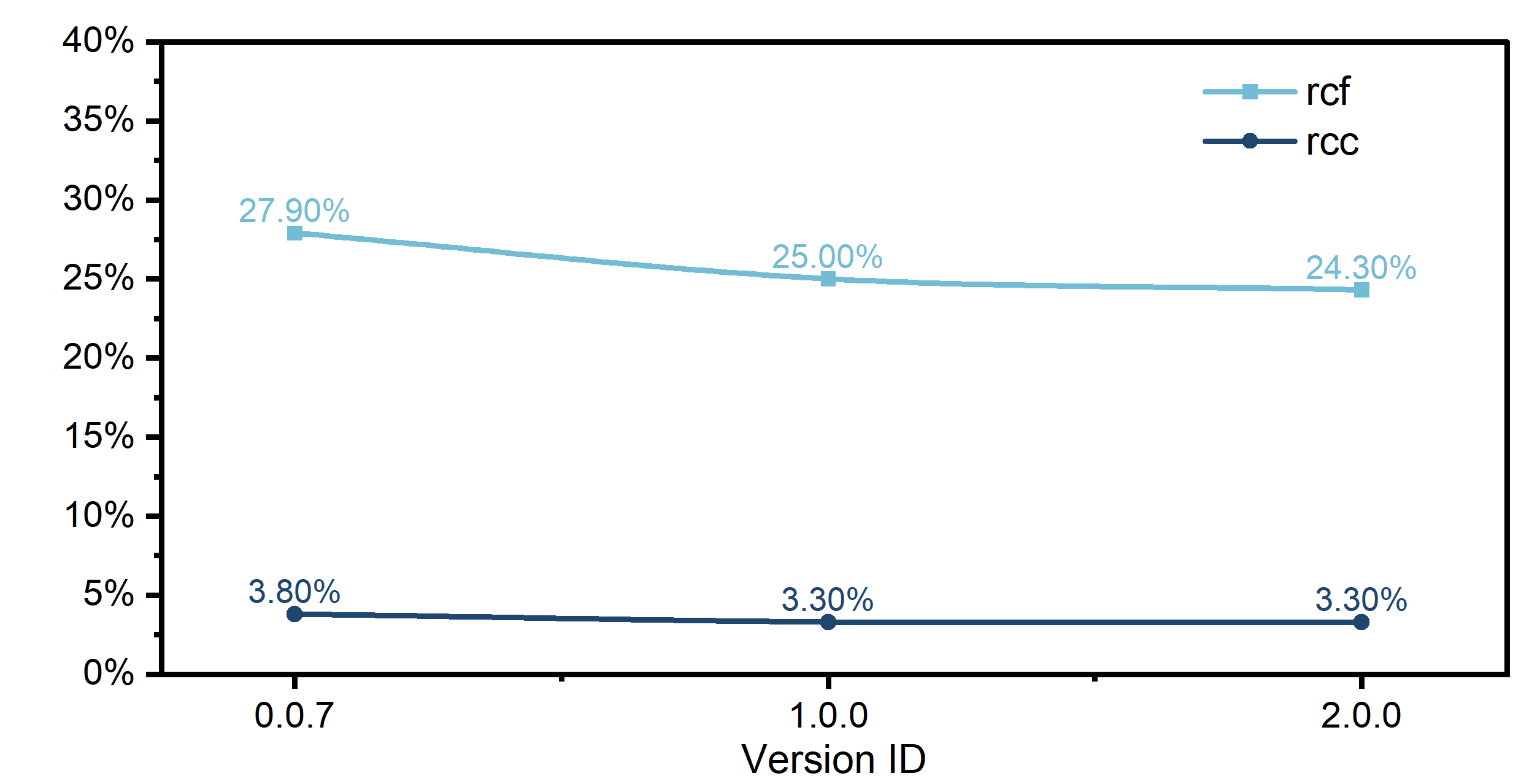}
        \subcaption{gpac}
        \label{fig:intra_gpac}
    \end{minipage}
    \hspace{0.01\columnwidth}
    \begin{minipage}{0.47\columnwidth}
        \centering
        \includegraphics[width=\columnwidth]{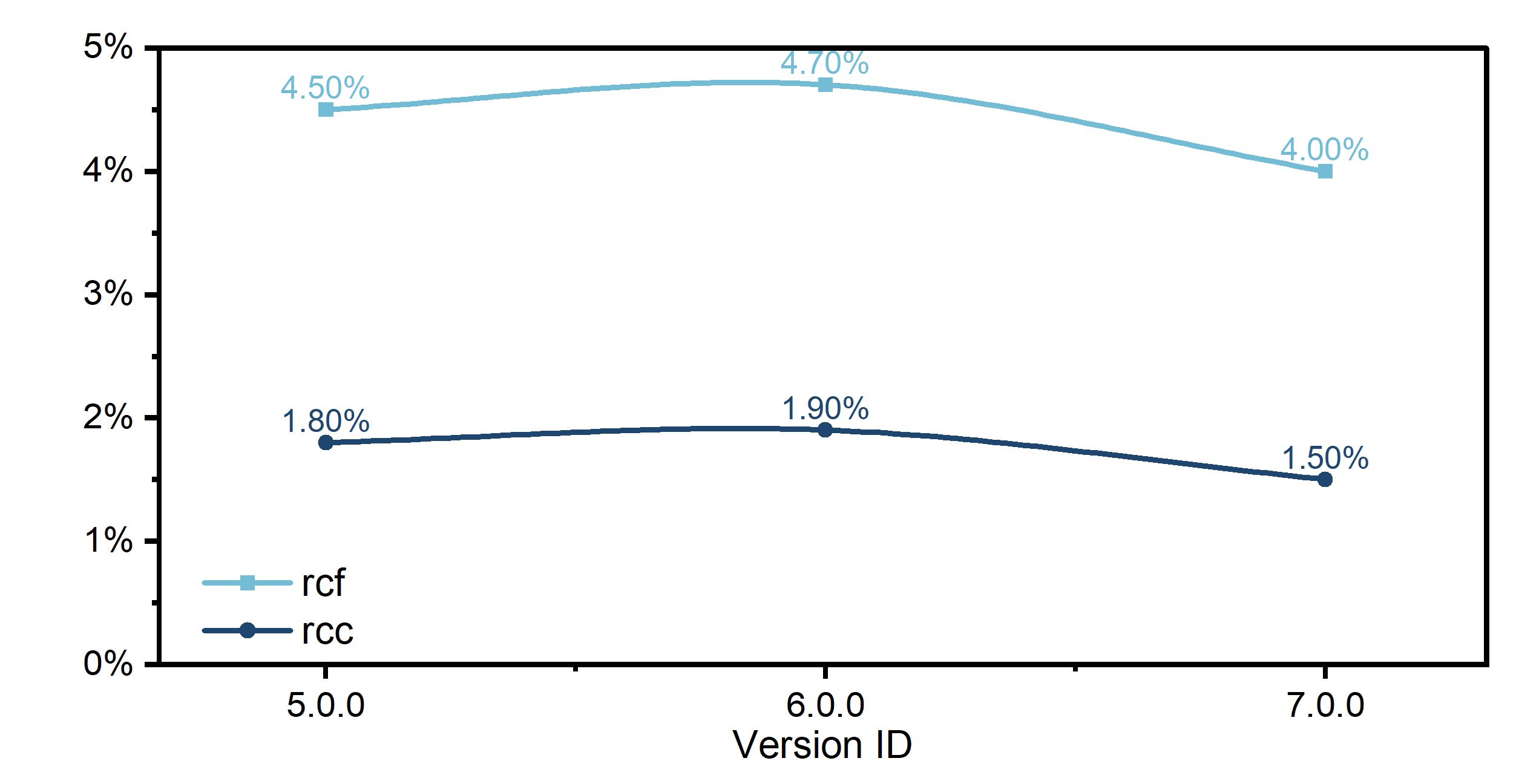}
        \subcaption{UnityPlugin}
        \label{fig:intra_up}
    \end{minipage}
    \caption{Detection results of intra-version clones across multiple versions when major version number changes.}
    \label{fig:intra_version_major}
\end{figure}

\begin{figure}[t]
    \centering
    \begin{minipage}{0.47\columnwidth}
        \centering
        \includegraphics[width=\columnwidth]{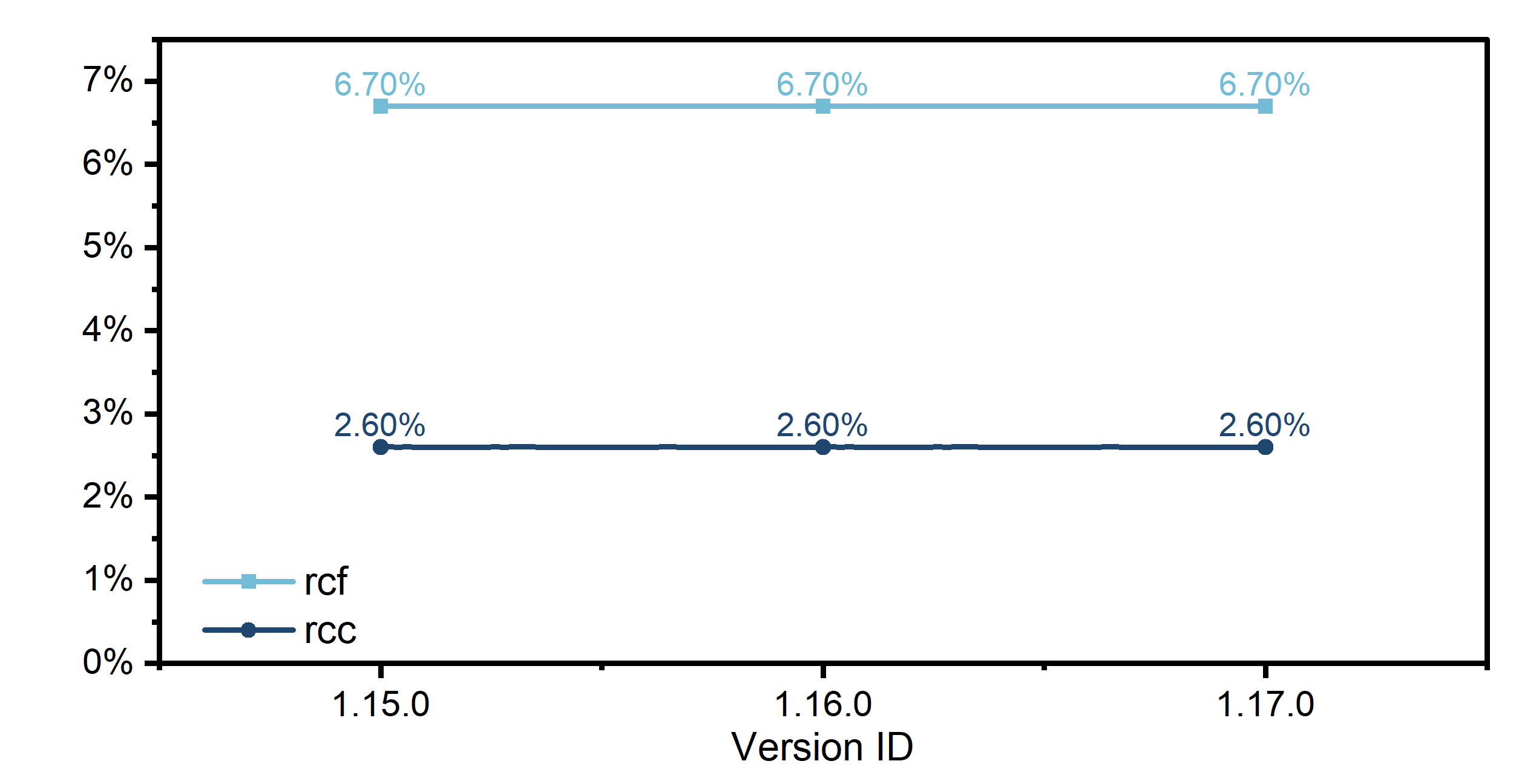}
        \subcaption{ViveInputUtility-Unity}
        \label{fig:intra_vivi}
    \end{minipage}
    \hspace{0.01\columnwidth}
    \begin{minipage}{0.47\columnwidth}
        \centering
        \includegraphics[width=\columnwidth]{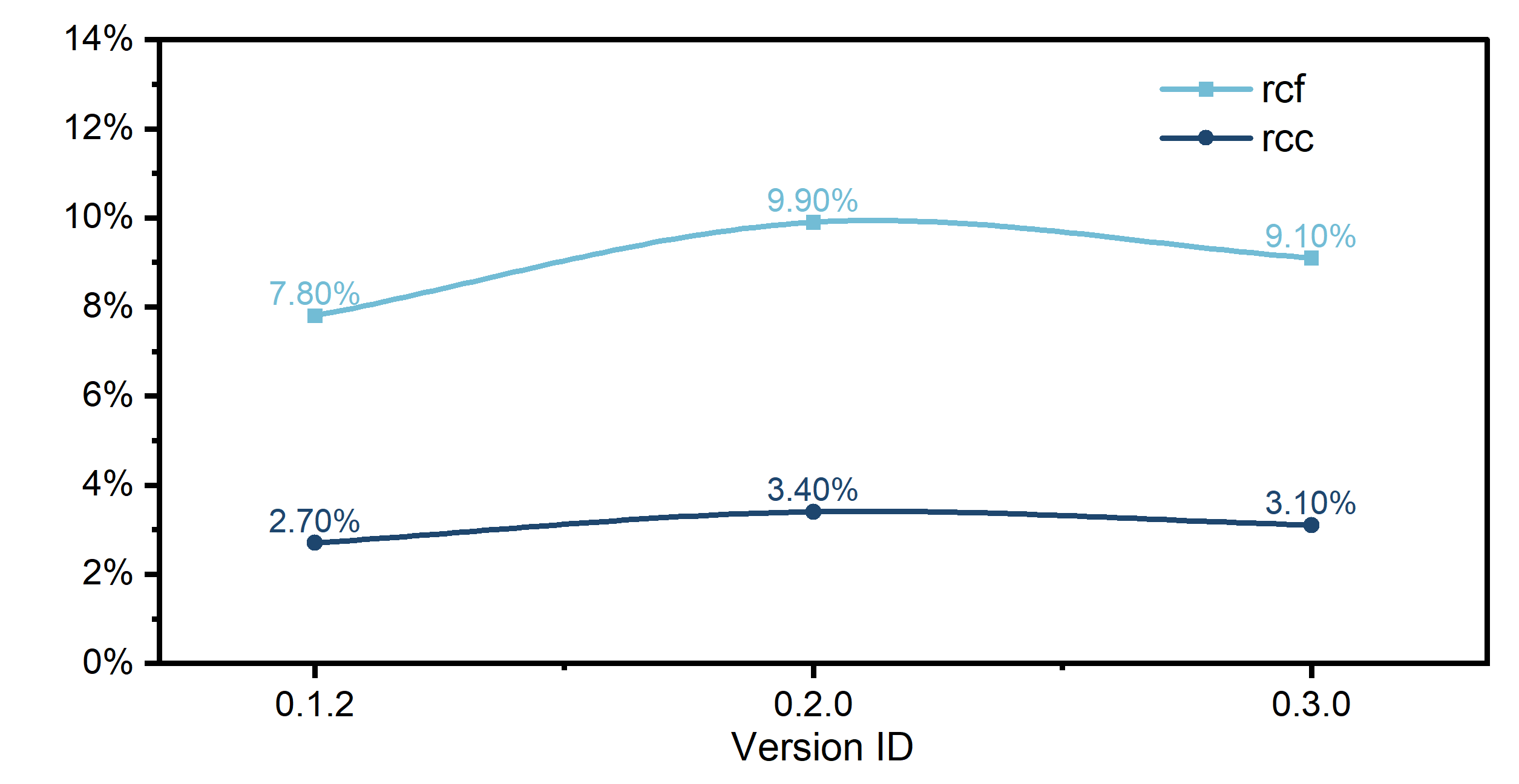}
        \subcaption{com.xrtk.core}
        \label{fig:intra_xrtk}
    \end{minipage}
    \caption{Detection results of intra-version clones across multiple versions when secondary version number changes.}
    \label{fig:intra_version_secondary}
\end{figure}

We find that in the ``open-brush" project, the $\rcc$ and $\rcf$ values fall within the ranges of [0.158, 0.163] and [0.348, 0.359], respectively. In the ``gpac" project, these values span wider ranges, with $\rcc$ between [0.168, 0.228] and $\rcf$ between [0.543, 0.664]. These observations suggest that while the degree of inter-version code cloning varies across projects, it remains consistently high, underscoring the prevalence of inter-version code clones in VR software development. This trend highlights that while code reuse improves development efficiency, it also poses the risk of propagating vulnerabilities, which may compromise software stability and security in future versions. 

To examine intra-version code clones, we divide the analysis into three sub-RQs focusing on their evolution with changes in (i) major version numbers (e.g., 0.x.x, 1.x.x, 2.x.x), (ii) secondary version numbers (e.g., 0.1.x, 0.2.x, 0.3.x), and (iii) ending version numbers (e.g., 0.1.1, 0.1.2, 0.1.3). For each sub-RQ, we select two distinct projects and analyze their latest three eligible versions. These projects are chosen for their frequent release schedules, which provide a robust dataset for analysis.

For {\em major version changes}, Figure \ref{fig:intra_gpac} and Figure \ref{fig:intra_up} illustrate how $\rcf$ and $\rcc$ evolve with major version changes in the VR projects ``gpac" and ``UnityPlugin", respectively. In the project ``gpac",  $\rcf$ fluctuates by up to 0.036 and $\rcc$ by up to 0.005 across three major versions. Similarly, in the project ``UnityPlugin", $\rcf$ shows a variation of up to 0.007, and $\rcc$ up to 0.004. These observations suggest that the extent of code clones, as measured by $\rcf$ and $\rcc$, remains relatively stable during major version updates for VR projects.

For {\em secondary version changes}, Figure \ref{fig:intra_vivi} and Figure \ref{fig:intra_xrtk} show the evolution of $\rcf$ and $\rcc$ as the secondary version numbers change in the projects ``ViveInputUtility-Unity" and ``com.xrtk.core", respectively. We observe that in the project ``ViveInputUtility-Unity", $\rcf$ and $\rcc$ remain constant across the three analyzed versions. In the project ``com.xrtk.core", $\rcf$ fluctuates by up to 0.021, and $\rcc$ by up to 0.007. This indicates that the level of code clones generally remains stable, or even unchanged, during secondary version updates.

For {\em ending version changes}, we do not present figures because the variations in $\rcf$ and $\rcc$ across such updates are negligible. The minimal differences suggest that code clone patterns exhibit almost no significant evolution at this level of versioning.

\insightbox{
Inter-version code cloning is more common than intra-version code cloning, introducing significant security risks due to the potential propagation of vulnerabilities across multiple versions of the software.}
}

\subsection{Asset File Cloning Analysis}

{\color{black}
\subsubsection{RQ7: How do cloning practices vary across different types of asset files in VR software?}

To answer this RQ, we select four representative projects from distinct application categories and conduct clone detection on the serialized files within each project, with results illustrated in Figure \ref{fig:clone_index}.

We observe a considerable disparity between scene files and material files, with scene files typically exhibiting the lowest clone level, averaging $\CI \approx 0.1125$, while material files demonstrate the highest, with an average of $\CI \approx 1.05$. This is possibly due to the fact that VR scenes generally contain unique content and structures that are difficult to replicate or share. Conversely, material files are extensively reused across various objects and scenes for optimization and performance reasons. 

Additionally, we find that the clone level of prefab files is intermediate, with an average of $\CI \approx 0.375$, yet it shows substantial fluctuations. This may be attributed to variations in system design and resource reuse strategies across different projects, which are heavily influenced by the developer's style. Some projects may require complex, customized prefabs, while others rely on a large number of standardized prefabs that are heavily reused. The use of techniques like prefab variants and runtime instantiation in complex prefabs, analogous to object-oriented programming, can lead to a reduced $\CI$. In contrast, extensive reuse of standardized prefabs results in a higher $\CI$, facilitating development but introducing redundancy, as changes to attributes need to be applied to every file within the clone group.

\begin{figure}[t]
    \centering
    \includegraphics[width=0.85\linewidth]{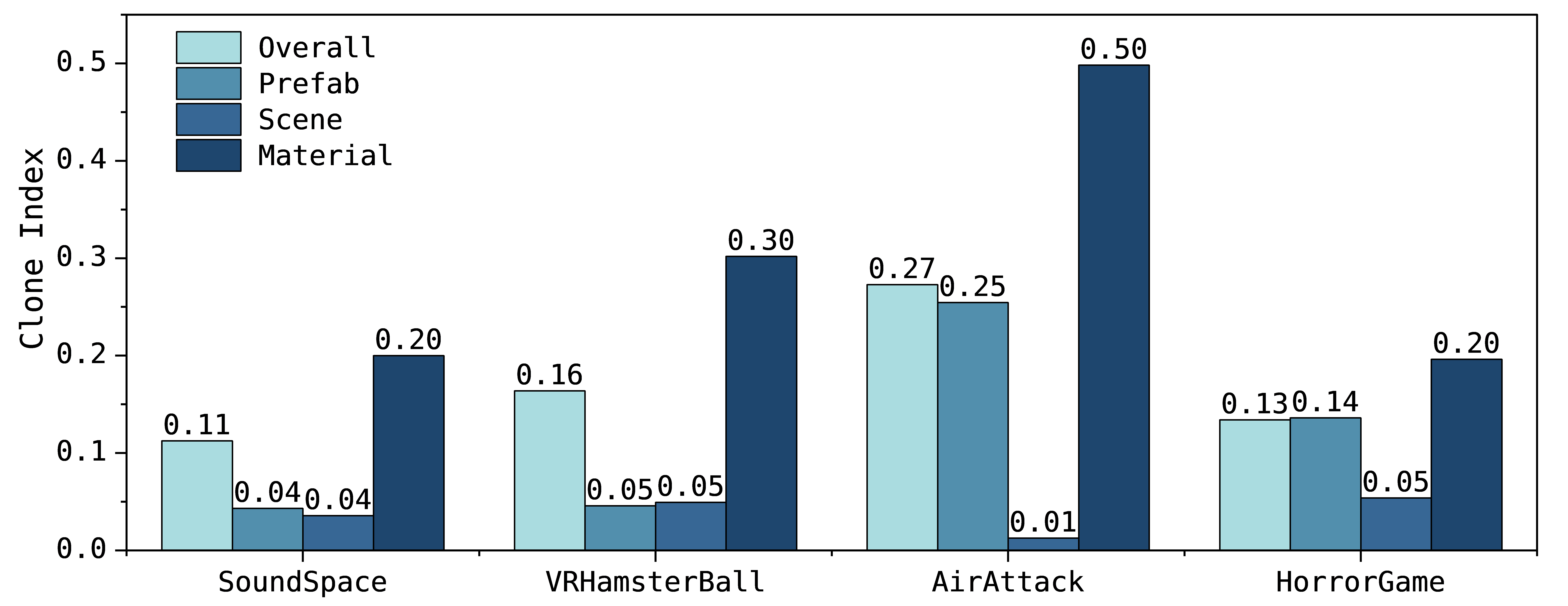}
    \caption{Analysis of cloning behavior of various asset files from four typical projects.}
    \label{fig:clone_index}
\end{figure}

\insightbox{In VR software, the cloning behavior of asset files varies with complexity. Complex files like scenes have lower clone levels, simpler ones like materials have higher, and prefab files fluctuate with development style.
}

\section{Discussion}
\label{sec:discussion}

\subsection{Recommendations}

\subsubsection{For Researchers} 
{This study yields several actionable implications for researchers studying software cloning in VR development. First, the strong correlation between project size and cloning suggests that scalable clone detection tools are necessary for large VR projects, particularly in the gaming and education domains where cloning is most prevalent. Second, the distinct reliance on serialized asset files rather than source code highlights the importance of developing specialized detection methods tailored to asset-based cloning. Third, security-focused research should particularly investigate inter-version cloning patterns that facilitate vulnerability propagation. Finally, to advance clone analysis in VR software, researchers should focus on creating context-aware, multi-faceted metrics that capture the unique structural and behavioral features of VR systems.  This effort should be supported by a comprehensive evaluation framework combining empirical validation, contextual experimentation, and cross-scenario applicability testing to ensure both precision and practical impact.}

\subsubsection{For VR Software Developers and Practitioners} 
{This study also yields several actionable implications for VR software developers and practitioners. First, employ robust design patterns and modular architectures to facilitate code reuse and minimize reliance on copy-paste practices, particularly in large-scale VR projects where code and asset cloning is more prevalent.

Second, the integration of clone detection tools into development workflows is essential for the early identification and refactoring of duplicated content, thereby mitigating long-term maintenance burdens in increasingly complex systems.

Third, asset management should be treated with the same rigor as source code maintenance. Rather than generating slight variations of materials or textures, developers are encouraged to establish a centralized asset repository and adopt reusable instances or prefabs to ensure consistency and reduce redundancy across scenes.

Fourth, development practices should align with the characteristics of the chosen programming language. In C\#, for instance, developers can leverage ScriptableObjects and inheritance to encapsulate reusable logic, whereas in C or C++, repetitive tasks should be delegated to optimized libraries or engine-level modules instead of being implemented repeatedly.

Fifth, evaluate third-party libraries carefully and consider alternatives with lower cloning levels to minimize clone. Developers should consider trimming or replacing libraries with high clone rates, especially when only a small subset of features is required. When robust frameworks are indispensable (e.g., Oculus SDK or SteamVR), their integration should be abstracted via dependency injection or interface wrappers to avoid deeply embedded code dependencies and facilitate easier maintenance.

Finally, to address the risks of inter-version cloning, teams should establish cross-version code management workflows. This includes centralizing reusable logic in shared core modules and maintaining traceability of duplicated segments to ensure timely propagation of fixes and enhancements across all project versions.
}

\subsection{Limitation and Future Work}
{The present study has several limitations that need to be addressed in future research. First, the scope of our dataset presents certain constraints that may affect the generalizability of our findings. (i) Currently, the dataset primarily consists of C\# projects, aligning with the current landscape of open-source VR development. Although the exclusion of non-NiCad-supported languages such as JavaScript affects only a small subset of auxiliary tools, extending language support is essential for enhancing generalizability—particularly in multilingual and proprietary VR contexts. (ii) Another key limitation of our study lies in the exclusion of Unreal Engine-based projects, despite their prominence in VR development. This exclusion stems from two primary factors: most Unreal Engine-based VR projects are proprietary and not publicly accessible as open-source code; and such projects are predominantly developed in C++, which is not supported by our clone detection tool, NiCad. These constraints collectively prevent the inclusion of Unreal Engine projects in our dataset and may limit the generalizability of our findings to closed-source VR ecosystems. We suggest that future research explore alternative analysis methods or tools capable of supporting C++ and binary formats, thereby broadening the scope of VR project analysis.

Second, the employment of GPT-4o for file clone detection presents several limitations that warrant discussion. (i) Different versions or configurations of the GPT-4o can sometimes produce varying similarity values, highlighting the need for consistent model configurations and fine-tuning to reduce discrepancies in similarity tasks. (ii) Although our study illustrates the effectiveness of GPT-4o in asset file clone detection for open-source VR software, its scalability and real-time performance in large-scale VR systems have not yet been evaluated, and current LLMs often underperform in such settings. In addition, while fine-tuned LLMs could improve detection precision for serialized assets, the absence of domain-specific datasets currently hinders such efforts. To address these issues, our future work will focus on building dedicated asset file corpora and training customized models. (iii) The application of LLMs in code and file clone detection within VR software introduces ethical concerns, particularly regarding data privacy and inference bias. While our reliance on open-source datasets mitigates these concerns in the present study, future research targeting proprietary VR projects or user-contributed content should adopt more robust privacy-preserving and fairness-aware practices.

Another limitation of this study lies in its absence of VR-specific clone metrics. The adopted metrics are derived from traditional software research and remain overly straightforward. Given the complexity and unique characteristics of VR environments, there exists a pressing need to establish more sophisticated, VR context-aware metrics that correspond to varied analytical objectives, coupled with systematic evaluations encompassing strengths, limitations, verification, and scenario-specific applicability.

Finally, our current evaluation methodology examines source code and serialized asset cloning in isolation. An important research direction would involve developing an integrated assessment framework that simultaneously analyzes both artifact types. This may be achieved through the design of a specialized LLM-based analysis agent capable of cross-artifact pattern recognition, unified clone behavior modeling, and combined metric computation—thereby enabling truly holistic VR software clone analysis.
}
\section{Related Work}
\label{sec:related_work}

\subsection{Empirical Study of VR Software}
The empirical study of VR software has gained increased attention in recent years, and the scope includes educational \cite{Oyelere2020Exploringa}, marketing communication \cite{Grudzewski2018Virtuala},  user operational performance \cite{Epp2021empirical}, automated testing \cite{Rzig2023Virtual}, vulnerabilities \cite{rodriguez2017empirical, Dastgerdy2024Virtual, Guo2024Empirical} and code cloning \cite{huang2024study}. Among the studies, Our prior work \cite{huang2024study} provides the most relevant research on code cloning in open-source VR software. However, the study does not address the code cloning of serialized files, which represents an important aspect of VR software.

\subsection{Code Clone Detection}

Traditional methods for detecting code clones include text-based, token-based, and AST-based techniques. Tools like Simian \cite{simian}, CCfinder \cite{1019480,kamiya2002ccfinder}, and NiCad \cite{NiCadClo0:online,cordy2011nicad} focus on comparing code at the syntactic or token level, while others like CloneDR \cite{CloneDR:online,OCallahan2003CloneDR} and Duplication Finder \cite{DuplicationFinder:online,Garcia2009DuplicationFinder} leverage the Abstract Syntax Tree for deeper analysis of code structure.

Recent advancements have introduced the use of Large Language Models (LLMs), such as Codex \cite{Codex:online,Chen2021Codex} and CodeBERT \cite{CodeBERT:online,Feng2020CodeBERT}, to detect clones with semantic awareness. These models understand code functionality, allowing them to identify clones that may not be syntactically identical but are semantically equivalent. By leveraging LLMs, clone detection can move beyond superficial matches, improving accuracy in detecting complex clones and offering greater flexibility across different programming languages. Recent research \cite{dou2023understandingcapabilitylargelanguage} shows that NiCad performs best in non-LLMs-based detection, while GPT-4 performs best in LLMs-based detection. Both of them achieve the highest precision and recall for each clone type in the same type of comparison.

\subsection{File Clone Detection}
Asset files in VR software, representing the differences from traditional software, can be considered specialized text files. There are several methods to detect their similarity, with some of the more representative ones being content-based methods, fingerprint-based methods, and semantic (deep learning) methods.

Content-based detection methods \cite{salton1983introduction,jones1972statistical,levenshtein1966binary,jaccard1901distribution} focus on directly comparing the file contents using traditional techniques such as string matching and text block comparison. However, these methods tend to be less efficient when dealing with large-scale systems. Fingerprint-based detection methods \cite{10.14778/2850469.2850470,rivest1992md5} improve efficiency by generating fingerprints for the files and using efficient lookup algorithms, making them particularly suitable for large-scale source code repositories. However, they cannot detect clones caused by syntax or structural changes. Deep learning-based detection methods \cite{Pennington2014GloVeGV,bojanowski2017enrichingwordvectorssubword,peters2018deepcontextualizedwordrepresentations,devlin2019bertpretrainingdeepbidirectional} rely on neural networks to model the semantics of the code or text. These methods can identify clones that differ in structure but are semantically similar, offering strong adaptability and high accuracy. However, they come with greater implementation complexity and computational overhead. Nevertheless, with the recent development of LLM technology and architecture, the disadvantages of deep learning-based detection methods have been alleviated to a certain extent, making it an advantageous method for similarity detection of serialized text files.

\section{Conclusion}
\label{sec:conclusion}
In this paper, we conduct a quantitative empirical study on code cloning in open-source VR projects. 
We refine the VR-345 dataset from previous studies for this research, ensuring a balance between the influence and range of the projects. We then propose a set of metrics to measure code cloning from different perspectives. Finally, we perform empirical experiments on seven carefully designed research questions at three distinct levels, in order to capture an overview of code cloning in VR software.

Our study reveals the key differences between VR software and traditional software in terms of code cloning. Due to these differences, certain files cannot be detected by traditional detection tools, which is why we employ large language model technology to assist in our detection. We evaluate and analyze the experimental results from various aspects such as clone distribution, clone concentration, programming languages, third-party libraries, and the impact of different types of assets. Through this, we answer the seven proposed research questions, deepening our understanding of code cloning in VR software.

\bibliography{main}

\end{document}